\documentclass[11pt, titlepage]{article}
\usepackage[sort&compress, round, semicolon, authoryear]{natbib}
\usepackage{graphicx, lscape}
\usepackage{graphics, color}
\usepackage{epsfig}
\usepackage{epstopdf}
\usepackage{wrapfig}
\usepackage{caption}
\usepackage{setspace}
\usepackage{subcaption}
\usepackage{amsmath, amsthm, amssymb, amscd}
\usepackage{latexsym}
\usepackage{multirow} 
\usepackage{array}
\usepackage{color}
\usepackage{rotating}
\usepackage[hyphens]{url}
\usepackage[hidelinks]{hyperref}  
\usepackage{cleveref}
\usepackage{fullpage}
\usepackage{fancyvrb}
\usepackage{pdfpages}
\usepackage{fmtcount}
\usepackage[margin=3.3cm]{geometry}
\usepackage{verbatim}
\usepackage[english]{babel}
\usepackage[utf8]{inputenc}
\usepackage{tikz} 
\usetikzlibrary{arrows, decorations.pathmorphing, backgrounds, fit, positioning, shapes.symbols, chains}
\usetikzlibrary{decorations.markings}
\usepackage{lscape}
\usepackage{algpseudocode}
\usepackage{algorithm}
\usepackage{soul}
\usepackage{booktabs}
\usepackage{enumerate}
\usepackage{colortbl}
\long\def\comment#1{} 
\usepackage{booktabs}
\usepackage{yhmath}  
\usepackage{mathtools}
\usepackage{rotating}
\usepackage{bm,upgreek}
\usepackage{mathrsfs}
\usepackage{changepage}
\usepackage{enumitem}
\usepackage{comment}
\usepackage[title,titletoc]{appendix}
\usepackage{kotex}

\usepackage{xcolor}
\definecolor{jm}{rgb}{0.5, 0.1, 0.3}
\definecolor{jh}{rgb}{0.2, 0.1, 0.6}
\definecolor{dark_green}{rgb}{0.1, 0.5, 0.1}

\newcommand{\romnum}[1]{\lowercase\expandafter{\romannumeral #1\relax}}

\newtheorem{proposition}{Proposition}[section]
\newtheorem{corollary}{Corollary}[section]
\newtheorem{theorem}{Theorem}[section]

\theoremstyle{definition}
\newtheorem{definition}{Definition}[section]

\usepackage{mathtools}

\newcommand{\bdm}{\begin{displaymath}}
	\newcommand{\edm}{\end{displaymath}}

\newcommand{\bs}{\boldsymbol}
\newcommand{\mc}{\mathcal}
\newcommand{\mr}{\mathrm}

\begin{document}
\newpage
\begin{center}
{\Large Principal Component Analysis in the Graph Frequency Domain
\medskip
}
\vskip 7mm

{\sc Kyusoon Kim and Hee-Seok Oh}\\
{Seoul National University, Seoul 08826, Korea}
\vskip 5mm

\today
\end{center}
\vskip 5mm
\begin{quote}
\begin{center}
\textbf{Abstract}
\end{center}

We propose a novel principal component analysis in the graph frequency domain for dimension reduction of multivariate data residing on graphs. The proposed method not only effectively reduces the dimensionality of multivariate graph signals, but also provides a closed-form reconstruction of the original data. In addition, we investigate several propositions related to principal components and the reconstruction errors, and introduce a graph spectral envelope that aids in identifying common graph frequencies in multivariate graph signals. We demonstrate the validity of the proposed method through a simulation study and further analyze the boarding and alighting patterns of Seoul Metropolitan Subway passengers using the proposed method.

\end{quote}	
	
	\pagenumbering{arabic}
	
\section{Introduction} \label{sec:intro}
Advances in monitoring tools, data collection, and storage devices have increased the number of observations and the dimensionality of data, i.e., the number of variables associated with each observation. Analyzing high-dimensional data as-is without dimensionality reduction can lead to the term ``curse of dimensionality," coined by \citet{Bellman1961}, and can pose challenges such as overfitting, poor accuracy of classification and prediction, and difficulty in visualization \citep{Ji2012big, Johnstone2009statistical, Donoho2000high}. As a result, researchers from several fields, such as statistics, signal processing, and machine learning, have focused on dimension reduction techniques. These methods aim to find low-dimensional data representations that not only save resources by reducing dimensions but also extract and summarize meaningful information from data. 

There are many well-known dimension reduction methods in Euclidean space, including principal component analysis (PCA), factor analysis, and independent component analysis (ICA). These methods have been extended to develop some dimensionality reduction techniques that can be applied to different data domains, taking into account the intrinsic characteristics of each domain. For example, functional PCA and functional ICA have been developed for functional data \citep{Shang2014survey, Hall2010principal, Virta2020independent, Pena2014}. Several studies have proposed dimension reduction methods on manifolds, such as symmetric positive definite matrices \citep{Harandi2017dimensionality} and spheres \citep{Lee2020spherical}. Dimension reduction techniques have also been explored for multivariate time series data in the time domain \citep{Pena2006dimension, Huang2010thesis, Park2010dimension, Lam2012factor, Egri2017cross, Lee2018martingale} and the frequency domain \citep{Shumway2017time, Brillinger2001time}.

The focus of this study is on the dimension reduction of multivariate data that resides on graphs. Graphs (or networks) are a powerful tool for describing the geometric structure of data, composed of vertices (or nodes) and edges. Data on graphs represent real (or complex) scalar values associated with each vertex, referred to as a graph signal. When the values at each vertex are real (or complex) vectors, this is called a multivariate graph signal. Examples of such data can be found around us, including measurements from sensor networks, traffic flows on road networks, epidemiological data on transportation networks, and signals from brain networks \citep{Shuman2013emerging}. When reducing the dimension of data on graphs, it is essential to consider the inherent graph structures while preserving as much of the information in the original data as possible. Therefore, dimension reduction methods specifically adapted to graph signals are necessary to effectively condense the information of multivariate graph signals.

We propose a new dimensionality reduction method, named {\em principal component analysis in the graph frequency domain}, explicitly designed for weakly stationary graph signals. While the previous studies reviewed in Section A of the supplementary material mainly focus on the vertex domain, our method operates in the graph frequency domain. To the best of our knowledge, this paper presents the first attempt to develop and implement a PCA method specifically in the graph frequency domain. To be specific, the proposed approach offers three main contributions. First, it effectively incorporates the underlying graph structures. Second, it identifies common graph frequencies in multivariate graph signals using the suggested graph spectral envelope and simultaneously estimates the normalized amplitudes (loading values) of the bases (eigenvectors) of the graph signal for each dimension. Finally, it provides a closed-form reconstruction to recover the original data from the dimension-reduced data. To demonstrate the effectiveness of the proposed method, we conduct a simulation study on various graphs and further apply the proposed method to Seoul Metropolitan subway passenger data to analyze the boarding and alighting patterns across different time intervals.

The rest of this paper is organized as follows. \Cref{sec:background} reviews principal component analysis in the frequency domain for multivariate time series, which inspired our method, along with basic concepts of graph signal processing. In \Cref{sec:gfreqPCA}, we present the proposed method with several key propositions. Numerical experiments, including simulation study and real data analysis, are discussed in  \Cref{sec:numericalexp}. Concluding remarks are given in \Cref{sec:conclude}. 
The {\tt R} codes used for the numerical experiments are available at \url{https://github.com/qsoon/gFreqPCA}.

\section{Background} \label{sec:background}
\subsection{PCA in the Frequency Domain} \label{sec:freqPCA}
We briefly review PCA in the frequency domain, a dimension reduction technique for multivariate time series. Suppose we have a $p$-dimensional stationary time series $Z(t)=(Z_1(t), \ldots, Z_p(t))^\top$. To reduce the dimensionality of $Z(t)$, one might consider applying classical PCA by performing an eigendecomposition of the $p\times p$ covariance matrix $\Sigma_{ZZ}$. However, this approach assumes that observations at each time point are independent, ignoring the temporal dependencies inherent in time series data, which is inappropriate for such an analysis. A more suitable alternative is to perform principal component analysis in the frequency domain, as proposed by \citet{Brillinger2001time} and detailed below.

Let $\mu_Z=E[Z(t)]$ be the mean of the time series, and let $\gamma_{ZZ}(u) = E[(Z(t+u)-\mu_Z)(Z(t)-\mu_Z)^H]$ be its autocovariance function, where $(Z(t)-\mu_Z)^H$ denotes the conjugate transpose of $(Z(t)-\mu_Z)$. The $p \times p$ power spectral density matrix, $f_{ZZ}$, is defined as $f_{ZZ}(\omega) = \frac{1}{2\pi}\sum_{u=-\infty}^{\infty} \gamma_{ZZ}(u) e^{-i\omega u}$. We obtain a dimension-reduced $q$-dimensional series $\zeta(t)$ using a $q \times p$ filter $b(t)$ as
\begin{equation} \label{eq:freqPCA_reduce}
    \zeta(t) = b(t) * Z(t) = \sum_u b(t-u) Z(u),  
\end{equation}
where $*$ denotes convolution. Then, by using a $p \times q$ filter $c(t)$, we reconstruct time series $\hat{Z}(t)$ as 
\begin{equation} \label{eq:freqPCA_reconstruct}
    \hat{Z}(t)= \mu + c(t)*\zeta(t) = \mu + \sum_{u} c(t-u)\zeta(u). 
\end{equation} 
The goal is to minimize the mean squared error, $E[(Z(t)-\hat{Z}(t))^H(Z(t)-\hat{Z}(t))]$. 

\begin{theorem} \label{thm:brillinger}
		(\citealp{Brillinger2001time}, Theorem 9.3.1) For a $p$-dimensional stationary time series $Z(t)$ with mean $\mu_Z$ and  power spectral density matrix $f_{ZZ}(\omega)$, the time series $\hat{Z}(t)$ that minimizes the mean squared error $E[(Z(t)-\hat{Z}(t))^H(Z(t)-\hat{Z}(t))]$ is given by 
        \begin{equation*}
            \hat{Z}(t)= \hat{\mu} + \sum_{u} \hat{c}(t-u)\hat{\zeta}(u) = \hat{\mu} + \sum_{u} \hat{c}(t-u)\sum_v \hat{b}(u-v) Z(v),
        \end{equation*}
        where $\hat{\mu}$, $\hat{b}(t)$, and $\hat{c}(t)$ are defined as
        \begin{equation*}
            \begin{gathered}
                \hat{\mu} = \mu_Z - \left(\sum_t \hat{c}(t)\right)\left(\sum_t \hat{b}(t)\right)\mu_Z, \\
                \hat{b}(t) = \frac{1}{2\pi} \int_{0}^{2\pi} \hat{B}(\omega) e^{i\omega t}d\omega, \\
                \hat{c}(t) = \frac{1}{2\pi} \int_{0}^{2\pi} \hat{C}(\omega) e^{i\omega t}d\omega.
            \end{gathered}
        \end{equation*}
        The frequency responses of the filters are 
        \begin{equation*}
                \hat{C}(\omega) = (
					u_1(\omega) \mid  
					\cdots \mid
					u_q(\omega)) = (\hat{B}(\omega))^H,
        \end{equation*}
        where $u_i(\omega)$ is the $i$th eigenvector of $f_{ZZ}(\omega)$, corresponding to its $i$th eigenvalue of $\eta_i(\omega)$ (with $\eta_1(\omega) \ge \cdots \ge \eta_p(\omega) \ge 0$). The minimum mean squared error is obtained as $\int_{0}^{2\pi} \sum_{i>q} \eta_i(\omega) d\omega$.
\end{theorem}

Note that obtaining the column vectors of $\hat{C}(\omega)$ is equivalent to sequentially finding vectors $\beta(\omega)$ at each $\omega$ such that
\begin{equation*}
    \underset{\beta({\omega}) \neq \bs{0}}{\max} \frac{\beta(\omega)^H f_{ZZ}(\omega) \beta(\omega)}{\beta(\omega)^H\beta(\omega)},
\end{equation*}
subject to the constraint that they are orthogonal to each other. This process is analogous to classical PCA, where it sequentially maximizes $\frac{\beta^H \gamma_{ZZ}(0) \beta}{\beta^H\beta}$ for the covariance matrix $\gamma_{ZZ}(0)$.  The series $\zeta_i(t) = \sum_u \hat{b}_i^\top(t-u) Z(u)$ is referred to as the $i$th principal component series of $Z(t)$, where $\hat{b}_i^\top(t)$ is the $i$th row vector of $\hat{b}(t)$ as defined in the theorem above. The series $\zeta_i(t)$ has a power spectral density $\eta_i(\omega)$. Moreover, $\zeta_i(t)$ and $\zeta_j(t)$ ($i \neq j$) have zero coherency across all frequencies.

\subsection{Graph Signal Processing} \label{sec:gsp}
Let $\mc{G} = (\mc{V}, \mc{E}, W)$ be an undirected or a directed weighted graph with a set of $n$ vertices $\mc{V}$, a set of edges $\mc{E}$, and a weighted adjacency matrix $W$. The $(i,j)$th entry of $W$ represents the strength of the connection between vertex $i$ and vertex $j$, with a zero value indicating no connection. A (multivariate) graph signal is a function from $\mc{V}$ to $\mathbb{R}^N$ (or $\mathbb{C}^N$), representing data associated with the vertices of graph $\mc{G}$. When $N=1$, a univariate graph signal $x$ on $\mc{G}$ can be represented as a vector of length $n$, where each element corresponds to the signal value at each vertex. The graph shift operator (GSO), denoted by an $n \times n$ matrix $S$, is a local operator that maps the signal value at a specific vertex to a linear combination of the signal values at its neighboring vertices \citep{Sandryhaila2014, Segarra2018}. This operator is fundamental to graph signal processing because it encapsulates the local structure of graph $\mc{G}$ and is used in the graph Fourier transform (GFT) \citep{Sandryhaila2013GFT}. The most common example of a GSO is the graph Laplacian $L$, defined as $L = \operatorname{diag}(\{\sum_{j=1}^n w_{ij}\}_{i=1}^{n}) - W$, where $w_{ij}$ denotes the $(i,j)$th element of $W$. The GSO is typically assumed to be a normal matrix, allowing it to be expressed as $S = V \Lambda V^H$, where $\Lambda$ is a diagonal matrix with elements $\lambda_1, \ldots, \lambda_n$, $V$ is a unitary matrix, and $V^H$ denotes the conjugate transpose of $V = (v_1 \mid \cdots \mid v_n)$. 
    
For a graph signal $x \in \mathbb{R}^n$ (or $\mathbb{C}^n$), the GFT of $x$ is defined as $V^H x$, which transforms the signal from the vertex domain to the graph frequency domain (or graph spectral domain). Linear and shift-invariant graph filters are defined by polynomials in GSO $S$ \citep{Sandryhaila2013discrete}. Consequently, a linear and shift-invariant graph filter $\mr{H}$ can be expressed as $\mr{H} = \sum_{\ell=1}^{n} h_\ell S^{\ell-1} = V\operatorname{diag}(\{ h(\lambda_\ell)\}_{\ell=1}^n)V^H$, where $h(\lambda) = \sum_{\ell=1}^{n} h_\ell \lambda^{\ell-1}$ is the frequency response. In this paper, the term `graph filters' refers specifically to linear and shift-invariant graph filters without further clarification. The filtered output of a graph signal $x$ is then given by $\mr{H}x$. Graph filters play a crucial role in manipulating signals in the graph frequency domain, enabling applications such as graph frequency filtering, graph signal denoising, and signal reconstruction. The notations listed in Table \ref{tbl:notations} are used throughout this paper. 

\begin{table}[!t]
    \captionsetup{justification=justified}
    \caption{Notation list}
    \centering
    \renewcommand{\arraystretch}{1.4}
    \resizebox{\textwidth}{!}{ 
    \begin{tabular}{cc}
    \hline
    Symbol & Description \\ \hline
    $\mc{G} = (\mc{V}, \mc{E}, W)$ & Weighted graph $\mc{G}$ with vertices $\mc{V}$, edges $\mc{E}$, and  weighted adjacency matrix $W$ \\
    $n$ & The number of vertices in graph $\mc{G}$ \\
    $S$ & Graph shift operator (GSO) \\
    $L$ & Graph Laplacian \\
    $\{\lambda_\ell \}_{\ell=1}^{n}$ & Graph frequencies \\
    $V$ & Matrix of eigenvectors $v_i$ of $S$ (or $L$) corresponding to $\lambda_i$\\
    $\Lambda$ & Diagonal matrix of $\lambda_i$ \\
    $X$ & A $p$-dimensional graph signal consisting of $X_1, \ldots, X_p$ \\
    $Y$ & A $q$-dimensional dimension-reduced graph signal consisting of $Y_1, \ldots, Y_q$ \\
    $Y_i$ & The $i$th principal component graph signal \\
    $\hat{X}_i$ & Reconstructed graph signal of $X_i$ \\
    $\Sigma_{ij}^X$ & Cross-covariance between the two univariate graph signals $X_i$ and $X_j$ that comprise $X$ \\
    $p_{ij}^X$ & Graph cross-spectral density (or power spectral density) of $X_i$ and $X_j$ \\
    $P_X(\lambda_\ell)$ & Spectral matrix with the $(i, j)$th element $p_{ij}^X(\lambda_\ell)$ \\
    $U(\lambda_\ell)$ & Matrix of eigenvectors $u_i(\lambda_\ell)$ of $P_X(\lambda_\ell)$ \\
    $\mr{T}(\lambda_\ell)$ & Diagonal matrix of eigenvalues $\tau_i(\lambda_\ell)$ of $P_X(\lambda_\ell)$ \\
    $V^H$ & Conjugate transpose of $V$ \\
    $v^*$ & Elementwise complex conjugate of $v$ \\
    $\lVert \cdot \rVert_2$ & $L_2$ norm \\
    $\lVert \cdot \rVert_1$ & $L_1$ norm \\
    \hline
    \end{tabular}
    \label{tbl:notations}
    }
\end{table}

\subsection{Stationarity on Graphs} \label{sec:graph_stationarity}
The notion of weak stationarity for univariate random graph processes has been explored in several studies \citep{Girault2015thesis, Girault2015, Perraudin2017, Marques2017}. In this study, we adopt the definition provided by \citet{Marques2017}.

\begin{definition} \label{def:weakstationary}
		\citep{Marques2017} Given a normal graph shift operator $S$, a zero-mean univariate random graph process $X$ is weakly stationary with respect to $S$ if the covariance matrix $\Sigma_{X} = E\left[X X^H\right]$ and $S$ are simultaneously diagonalizable.
	\end{definition}

If $X$ is not a zero-mean process, the first moment of $X$ is required to be $E[X] \propto v_i$ for $1 \le i \le n$. 

\citet{Kim2024cross} defined the concept of joint weak stationarity of bivariate graph process as follows.
\begin{definition} \label{def:jointweakstationary}
		\citep{Kim2024cross} Given a normal graph shift operator $S=V \Lambda V^H$, zero-mean univariate random graph processes $X$ and $Y$ are jointly weakly stationary with respect to $S$ if (i) they are both weakly stationary with respect to $S$ and (ii) the cross-covariance matrix $\Sigma_{XY} = E[X Y^H]$ and $S$ are simultaneously diagonalizable.
	\end{definition}
If $X$ and $Y$ are not zero-mean processes, the first moments of $X$ and $Y$ are required to be $E[X] \propto v_i$ and $E[Y] \propto v_j$ for $1 \le i, j \le n$. \citet{Kim2024cross} also defined the graph cross-spectral density (GCSD) and the graph coherence as follows.

\begin{definition} \label{def:gcsd}
		\citep{Kim2024cross} Given a normal graph shift operator $S=V \Lambda V^H$, let $X$ and $Y$ be zero-mean, jointly weakly stationary univariate random graph processes with respect to $S$. The graph cross-spectral density of $X$ and $Y$ is defined as an $n \times 1$ vector $p_{XY} = \operatorname{diag}(V^H \Sigma_{XY} V)$, where $\Sigma_{XY} = E[X Y^H]$ denotes the cross-covariance matrix. 
		The coherence between $X$ and $Y$ is defined as $c_{XY} = \frac{\lvert p_{XY} \rvert^2}{p_{XX} p_{YY}}$, where the product and division operators are applied elementwise.
	\end{definition}

Note that if $X=Y$, the definition of GCSD of $X$ and $Y$ simplifies to the definition of graph power spectral density (GPSD) as defined by \citet{Marques2017}. Refer to \citet{Kim2024cross} for more information.

We now define the weak stationarity of a multivariate random graph process.

\begin{definition} \label{def:multivariate_stationarity}
		Given a normal graph shift operator $S$, consider a zero-mean $p$-dimensional random graph process $X=(X_1 \mid \cdots \mid X_p)$, where each $X_i$ ($1 \le i \le p$) is an $n$-dimensional vector. The graph signal $X$ is said to be weakly stationary if the cross-covariance matrices $\Sigma_{ij}^X:= \operatorname{Cov}(X_i, X_j)$ and $S$ are simultaneously diagonalizable for all $1\le i, j \le p$.
	\end{definition}

If $X_i$ ($1 \le i \le p)$ are not zero-mean random graph processes, then their first moments are required to satisfy $E[X_i] \propto v_i$ ($1 \le i \le p)$. For simplicity, we will refer to the term `weak stationarity' as `stationarity' from now on. Moreover, for convenience, the term `graph signal' will be used to denote both the random graph process and its realization, depending on the context. The GPSD and GCSD can be estimated using the windowed average graph periodogram and the windowed average graph cross-periodogram, respectively. Random windows can be practically utilized for window design \citep{Kim2024cross}.

Before closing this section, we remark that the multivariate graph signal literature is no stranger to the concept of dimension reduction, and references are diverse. However, most studies have been developed in the vertex domain, which are overviewed in Section A of the supplementary material.

\section{PCA in the Graph Frequency Domain} \label{sec:gfreqPCA}
In this section, we propose a new PCA in the graph frequency domain. Consider a $p$-dimensional stationary graph signal $X = (X_1 \mid \cdots \mid X_p)$ with respect to GSO $S$ on $\mathcal{G}$, where each $X_i$ ($1 \le i \le p$) is an $n$-dimensional vector. We denote the mean of $X_i$ by $\mu_i^X = E[X_i]$ and the cross-covariance between $X_i$ and $X_j$ by $\Sigma_{ij}^X:= \operatorname{Cov}(X_i, X_j)$ for $1 \le i,j \le p$. Let $p_{ij}^X = (p_{ij}^X(\lambda_1), \ldots, p_{ij}^X(\lambda_n))^\top$ represent the GCSD. When $i=j$, $p_{ii}^X$ becomes the GPSD. If we define $P_X(\lambda_\ell)$ as a $p \times p$ matrix whose $(i,j)$th element is $p_{ij}^X(\lambda_\ell)$, the matrix $P_X(\lambda_\ell)$ is Hermitian and positive semi-definite by Proposition \ref{prop:psd_Plambda}, ensuring that its eigenvalues are real and nonnegative.

\begin{proposition} \label{prop:psd_Plambda}
    For a $p$-dimensional stationary graph signal $X=(X_1 \mid \cdots \mid X_p)$ with respect to the graph shift operator $S=V\Lambda V^H$ on $\mc{G}$, let $p_{ij}^X$ denote the graph cross-spectral density of $X_i$ and $X_j$. Then, a $p \times p$ matrix $P_X(\lambda_\ell)$ is Hermitian and positive semi-definite for each $\lambda_\ell$ ($1 \le \ell \le n$), where the $(i,j)$th element is $p_{ij}^X(\lambda_\ell)$.
\end{proposition}

A proof is provided in Section B.1 of the supplementary material. 

\begin{corollary} \label{cor:Plambda_decomp}
    For each $\lambda_\ell$ ($1 \le \ell \le n$), there exists a unitary matrix $U(\lambda_\ell)$ and a real diagonal matrix $\mr{T}(\lambda_\ell)$ such that $P_X(\lambda_\ell) = U(\lambda_\ell) \mr{T}(\lambda_\ell) U(\lambda_\ell)^H$, where the $i$th column of $U(\lambda_\ell)$ is $u_i(\lambda_\ell)$ and the $i$th diagonal element of $\mr{T}(\lambda_\ell)$ is $\tau_i(\lambda_\ell)$ with $\tau_1(\lambda_\ell) \ge \cdots \ge \tau_p(\lambda_\ell) \ge 0$.
\end{corollary}

We consider the problem of reducing the dimensionality of $X$ from $p$ to $q$ $(< p)$ using graph filters, which results in a dimension-reduced graph signal $Y=(Y_1 \mid \cdots \mid Y_q)$. Analogous to (\ref{eq:freqPCA_reduce}), each component $Y_i$ is defined as a sum of graph-filtered outputs of the signal, given by
\begin{equation} \label{eq:gfreqPCA_reduce}
    Y_i = \mr{H}_{i1}X_1 + \cdots + \mr{H}_{ip}X_p, \quad 1 \le i \le q,
\end{equation}
where $\mr{H}_{ij}=V \operatorname{diag}(\{h_{ij}(\lambda_\ell)\}_{\ell=1}^{n}) V^H$ are graph filters. Similar to (\ref{eq:freqPCA_reconstruct}), using graph filters $\mr{G}_{ij}=V \operatorname{diag}(\{g_{ij}(\lambda_\ell)\}_{\ell=1}^{n}) V^H$, we reconstruct the graph signal as
\begin{equation} \label{eq:gfreqPCA_reconstruct} 
    \hat{X}_i = \mu_i + \mr{G}_{i1}Y_1 + \cdots + \mr{G}_{iq}Y_q, \quad 1 \le i \le p,
\end{equation}
where $\mu_i$ is a vector of length $n$. We then obtain $\hat{X}_i = \mu_i+(\mr{G}_{i1} \mr{H}_{11} + \cdots + \mr{G}_{iq} \mr{H}_{q1})X_1 + \cdots + (\mr{G}_{i1} \mr{H}_{1p} + \cdots + \mr{G}_{iq} \mr{H}_{qp})X_p$, expressed as
\begin{equation} \label{eq:xhat_final}
    \hat{X}_i = \mu_i+\mr{A}_{i1}X_1+\cdots+\mr{A}_{ip}X_p, \quad 1 \le i \le p,
\end{equation}
where $\mr{A}_{ij}=\mr{G}_{i1} \mr{H}_{1j} + \cdots + \mr{G}_{iq} \mr{H}_{qj}$ is a graph filter with frequency responses $a_{ij}(\lambda_\ell)=g_{i1}(\lambda_\ell) h_{1j}(\lambda_\ell) + \cdots + g_{iq}(\lambda_\ell) h_{qj}(\lambda_\ell)$, i.e., $\mr{A}_{ij}=V \operatorname{diag}(\{a_{ij}(\lambda_\ell)\}_{\ell=1}^{n}) V^H$. We aim to minimize the following mean squared error, $\sum_{i=1}^{p} E[(X_i-\hat{X}_i)^H(X_i-\hat{X}_i)]$, where the solution is provided by the result below.

\begin{theorem} \label{thm:proposedsol}
    Suppose that $X=(X_1 \mid \cdots \mid X_p)$ be a $p$-dimensional stationary graph signal with respect to the graph shift operator $S=V\Lambda V^H$ on $\mc{G}$. Let $\mu_i^X$ be the mean of $X_i$, and $p_{ij}^X$ be the graph cross-spectral density of $X_i$ and $X_j$. For each graph frequency $\lambda_\ell$, define the $p \times p$ spectral matrix $P_X(\lambda_\ell)$ for $X$ whose $(i,j)$th element is $p_{ij}^X(\lambda_\ell)$. Then, the graph signal $\hat{X}_i$ defined by (\ref{eq:xhat_final}), which minimizes the mean squared error $\sum_{i=1}^{p} E[(X_i-\hat{X}_i)^H(X_i-\hat{X}_i)]$, is given by 
    \begin{equation*}
        \hat{X}_i= \hat{\mu}_i + \hat{\mr{A}}_{i1}X_1+\cdots+\hat{\mr{A}}_{ip}X_p, \quad 1 \le i \le p,
    \end{equation*}
    where $\hat{\mr{A}}_{ij}=V \operatorname{diag}(\{\hat{a}_{ij}(\lambda_\ell)\}_{\ell=1}^{n}) V^H$, 
    \begin{eqnarray*}
    \hat{\mu}_i &=& \mu_i^X -\sum_{j=1}^p \hat{\mr{G}}_{ij}(\sum_{k=1}^p \hat{\mr{H}}_{jk}\mu_k^X),\\
            \hat{A}(\lambda_\ell) &=& \hat{G}(\lambda_\ell) \hat{H}(\lambda_\ell), \\
            \hat{G}(\lambda_\ell) &=& (u_1(\lambda_\ell) | \cdots | u_q(\lambda_\ell))= (\hat{H}(\lambda_\ell))^H.
    \end{eqnarray*}
    Here, $\hat{\mr{H}}_{ij}=V \operatorname{diag}(\{\hat{h}_{ij}(\lambda_\ell)\}_{\ell=1}^{n}) V^H$ and $\hat{\mr{G}}_{ij}=V \operatorname{diag}(\{\hat{g}_{ij}(\lambda_\ell)\}_{\ell=1}^{n}) V^H$. The matrices $\hat{A}(\lambda_\ell)$, $\hat{H}(\lambda_\ell)$, and $\hat{G}(\lambda_\ell)$ have dimensions $p \times p$, $q \times p$, and $p \times q$, respectively, where the $(i,j)$th elements are $\hat{a}_{ij}(\lambda_\ell)$, $\hat{h}_{ij}(\lambda_\ell)$, and  $\hat{g}_{ij}(\lambda_\ell)$. $u_i(\lambda_\ell)$ denotes the $i$th eigenvector of $P_X(\lambda_\ell)$ corresponding to its $i$th eigenvalue, $\tau_i(\lambda_\ell)$ (with $\tau_1(\lambda_\ell) \ge \cdots \ge \tau_p(\lambda_\ell) \ge 0$). Note that the (minimum mean squared) reconstruction error is obtained as $\sum_{\ell=1}^{n} \sum_{i>q} \tau_i(\lambda_\ell)$.
\end{theorem}
A proof is provided in Section B.2 of the supplementary material.

Analogous to PCA in the frequency domain, obtaining the column vectors of $\hat{G}(\lambda_\ell)$ corresponds to sequentially determining vectors $\beta(\lambda_\ell)$ at each $\lambda_\ell$ such that 
\begin{equation} \label{eq:spectralenv}
    \underset{\beta(\lambda_\ell) \neq \bs{0}}{\max} \frac{\beta(\lambda_\ell)^H P_X(\lambda_\ell) \beta(\lambda_\ell)}{\beta(\lambda_\ell)^H \beta(\lambda_\ell)}
\end{equation}
subject to the condition that these vectors are orthogonal to each other. 

The components $Y_i = \sum_{k=1}^{p}\hat{\mr{H}}_{ik}X_k$ obtained in \Cref{thm:proposedsol} are called the {\em principal component graph signals}. The importance of a principal component graph signal can be measured by the percentage of the reduction in the reconstruction error attributable to that principal component graph signal. By \Cref{thm:proposedsol}, the reconstruction error is given by $\sum_{\ell=1}^{n} \sum_{i>q} \tau_i(\lambda_\ell)$ when using $q$ principal component graph signals. Thus, the fraction of the reconstruction error reduction attributable to the $i$th principal component graph signal $Y_i$ is $\sum_{\ell=1}^{n} \tau_i(\lambda_\ell) / \sum_{\ell=1}^{n} \sum_{j=1}^p \tau_j(\lambda_\ell)$. To determine the reduced dimension $q$, we generate a scree plot showing the proportion of the reconstruction error reduction as a function of $q$. The optimal $q$ corresponds to the elbow of the plot.

We evaluate the graph cross-spectral density between the principal component graph signals as follows.
\begin{proposition} \label{prop:zero_coherence}
    Under the conditions of Theorem \ref{thm:proposedsol}, the $i$th principal component graph signal $Y_i$ has a graph power spectral density given by $(\tau_i(\lambda_1), \ldots, \tau_i(\lambda_n))^\top$. Furthermore, the signals $Y_i$ and $Y_j$ ($i \neq j$) exhibit zero graph coherency across all graph frequencies.
\end{proposition}

A proof is provided in Section B.3 of the supplementary material. From \Cref{prop:zero_coherence}, the spectral matrix $P_Y(\lambda_\ell)$ for $Y = (Y_1 \mid \cdots \mid Y_q)$ is given by 
$
    P_Y(\lambda_\ell) = \operatorname{diag}(\{ \tau_i(\lambda_\ell)\}_{i=1}^q).
$

We can see that the GPSD of the first principal component graph signal $Y_1 = \sum_{k=1}^{p}\hat{\mr{H}}_{1k}X_k$ is the highest across all graph frequencies among the possible GPSDs of graph signals of the form $\sum_{k=1}^{p}\mr{H}_{1k}X_k$, given the constraint that $\lVert(h_{11}(\lambda_\ell), \ldots, h_{1p}(\lambda_\ell))^\top \rVert_2=1$. This can be inferred from the proof of \Cref{prop:zero_coherence}. In this context, we define the \textit{graph spectral envelope} as the maximum value of (\ref{eq:spectralenv}) for a given $\lambda_\ell$, 
\[
\tau(\lambda_\ell):=\underset{\beta(\lambda_\ell)}{\max} \frac{\beta(\lambda_\ell)^H P_X(\lambda_\ell) \beta(\lambda_\ell)}{\beta(\lambda_\ell)^H \beta(\lambda_\ell)},
\]
which is equal to both $\tau_1(\lambda_\ell)$ and the GPSD of the first principal component graph signal $Y_1$ at $\lambda_\ell$. The term \textit{envelope} indicates that the graph spectral envelope $\tau_1(\lambda_\ell)$ bounds all GPSDs of graph signals of the form $\sum_{k=1}^{p}\mr{H}_{1k}X_k$, under the constraint that $\lVert(h_{11}(\lambda_\ell), \ldots, h_{1p}(\lambda_\ell))^\top \rVert_2=1$. As a result, the graph spectral envelope $\tau_1(\lambda_\ell)$ should bound GPSDs of $X_1, \ldots, X_p$, making it a useful tool for detecting common graph frequencies present in the multivariate graph signal $X$. This is analogous to the results for the classical time series discussed in \citet{Stoffer1993spectral} and \citet{Stoffer1999detecting}.

Moreover, we define $\beta(\lambda_\ell)$, which optimizes (\ref{eq:spectralenv}) for each $\lambda_\ell$, as the \textit{optimal graph frequency scaling} at $\lambda_\ell$, which is equal to $u_1(\lambda_\ell)$. The optimal graph frequency scaling can be used to estimate the normalized amplitudes of the basis signals, the eigenvectors of GSO $S$. For example, suppose that $X_i = c_{ik} v_k + c_{im} v_m$ for $1 \le i \le p$, where $v_k$ and $v_m$ are the eigenvectors of GSO $S$ corresponding to the eigenvalues $\lambda_k$ and $\lambda_m$, respectively. Then, the GCSD $p_{ij}^X(\lambda_k) = c_{ik} c_{jk}^*$ for $1 \le i, j \le p$. Consequently, the spectral matrix $P_X(\lambda_k) = c_k c_k^H$, where $c_k = (c_{1k}, \ldots, c_{pk})^\top$. Therefore, the optimal graph frequency scaling at $\lambda_k$, $u_1(\lambda_k)$, is the unit norm solution $\beta(\lambda_k)$ that maximizes $\beta(\lambda_k)^H c_k c_k^H \beta(\lambda_k)$. By the Cauchy--Schwarz inequality, the solution is given by $u_1(\lambda_k) = c_k / \lVert c_k\rVert_2$ and similarly, $u_1(\lambda_m) = c_m / \lVert c_m\rVert_2$. Thus, the optimal graph frequency scaling at a certain graph frequency becomes the normalized amplitude of the basis signal corresponding to that graph frequency. Based on this observation, we can use the optimal graph frequency scaling to estimate the normalized amplitudes of the basis signals. 

By defining the error signal as $\epsilon_i = X_i - \hat{X}_i$ residing on the graph, we compute the graph cross-spectral density between the error signals $\epsilon_i$ ($1 \le i \le p$).

\begin{proposition} \label{prop:error_cpsd}
    Under the conditions of Theorem \ref{thm:proposedsol}, the error signal $\epsilon_i = X_i - \hat{X}_i$ has mean zero. Let $p_{ij}^\epsilon$ denote the GCSD between $\epsilon_i$ and $\epsilon_j$, and define the spectral matrix $P_\epsilon(\lambda_\ell)$ for $\epsilon = (\epsilon_1 \mid \cdots \mid \epsilon_p)$ at each graph frequency $\lambda_\ell$ whose $(i,j)$th element is $p_{ij}^\epsilon(\lambda_\ell)$. Then, we have  
    \begin{equation*}
        P_\epsilon(\lambda_\ell) = (I-\hat{\mr{A}}(\lambda_\ell)) P_X(\lambda_\ell) (I-\hat{\mr{A}}(\lambda_\ell))^H = \sum_{i>q} \tau_i(\lambda_\ell) u_{i}(\lambda_\ell) u_{i}^H(\lambda_\ell).
    \end{equation*}
\end{proposition}

\begin{proposition} \label{prop:xhat_coherence}
    Under the conditions of Theorem \ref{thm:proposedsol} and using the notations of \Cref{prop:error_cpsd}, the graph coherency between the $i$th principal component graph signal $Y_i$ and the error signal $\epsilon_j$ for all $1\le i,j \le p$ is zero across all graph frequencies. Also, the graph coherency between the error signal $\epsilon_j$ and the reconstructed graph signal $\hat{X}_k$ for all $1\le j,k \le p$ is zero across all graph frequencies. 
\end{proposition}

The proofs of \Cref{prop:error_cpsd} and \Cref{prop:xhat_coherence} are provided in Section B.4 and Section B.5 of the supplementary material, respectively.

As described, to implement the proposed method, named `gFreqPCA,' we need to estimate the GCSDs of the graph signals $X_i$ in $X$. For this purpose, we can use the graph cross-periodogram and the windowed average graph cross-periodogram proposed by \citet{Kim2024cross}.

\section{Numerical Experiments} \label{sec:numericalexp}
\subsection{Simulation Study} \label{sec:simul}
In this section, we perform a simulation study to investigate the ability of the graph spectral envelope to identify the common frequencies of a multivariate graph signal and the effectiveness of the optimal graph frequency scaling in estimating the normalized amplitudes of the basis signals (corresponding to the eigenvectors of GSO). 

\begin{figure}[!t]
    \centering
        \includegraphics[width=0.9\textwidth]{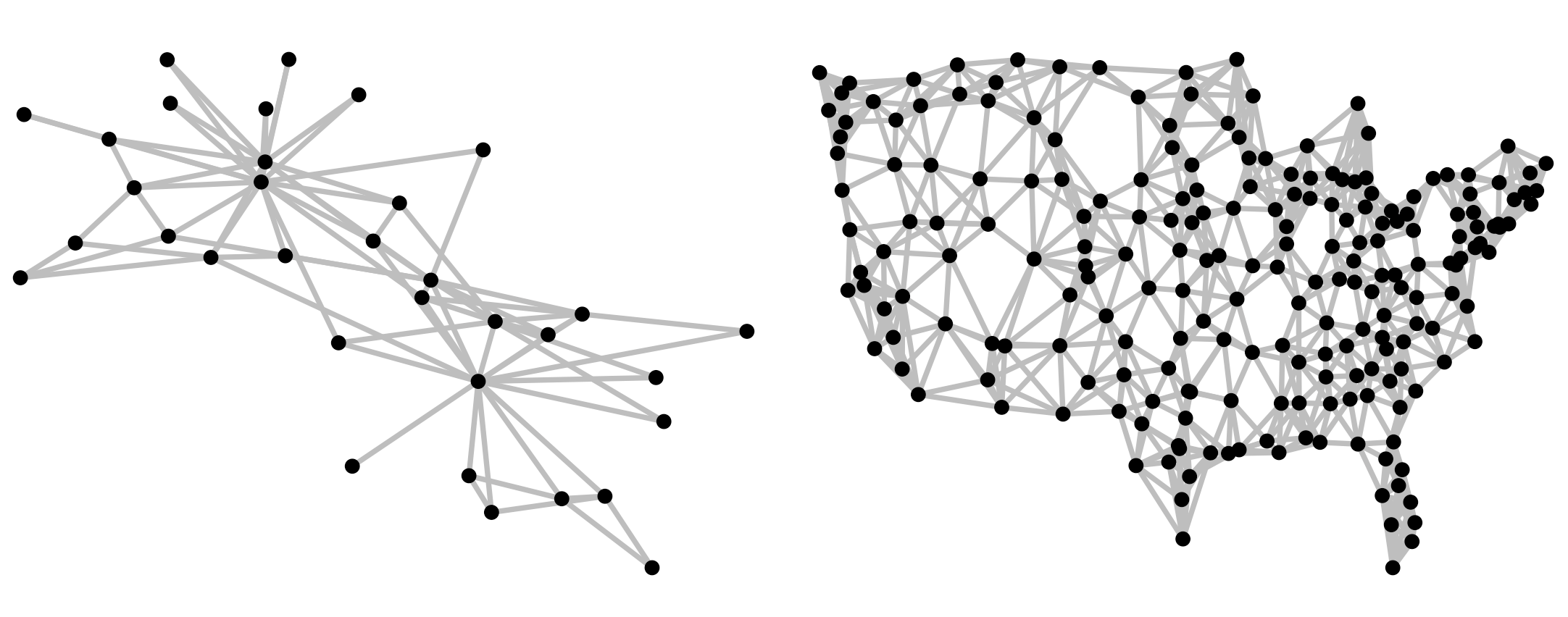}
        \vspace{-3mm}
    \caption{Two networks used in simulation study: Karate club network (left) and US sensor network (right).}
    \label{fig:simulgraphs}
\end{figure}

For the generation of graph signals, we consider two networks: the Karate club network \citep{Zachary1977}, consisting of 34 vertices and 78 edges, and the United States (US) sensor network \citep{Zeng2017bipartite}, shown in \Cref{fig:simulgraphs}. The US sensor network comprises 218 vertices, each connected to its seven nearest neighbors. The edge weights between vertex $i$ and vertex $j$ are defined as $w_{ij} = \exp(-d^2(i,j) / ave^2)$, where $d(i,j)$ is the distance between vertex $i$ and vertex $j$, and $ave$ denotes the average distance between all stations. For both cases, the graph Laplacian is used as GSO. The GPSD is estimated using the windowed average graph periodogram, calculated with 50 random windows.

\begin{figure}[!t]
    \centering
        \includegraphics[width=0.95\textwidth]{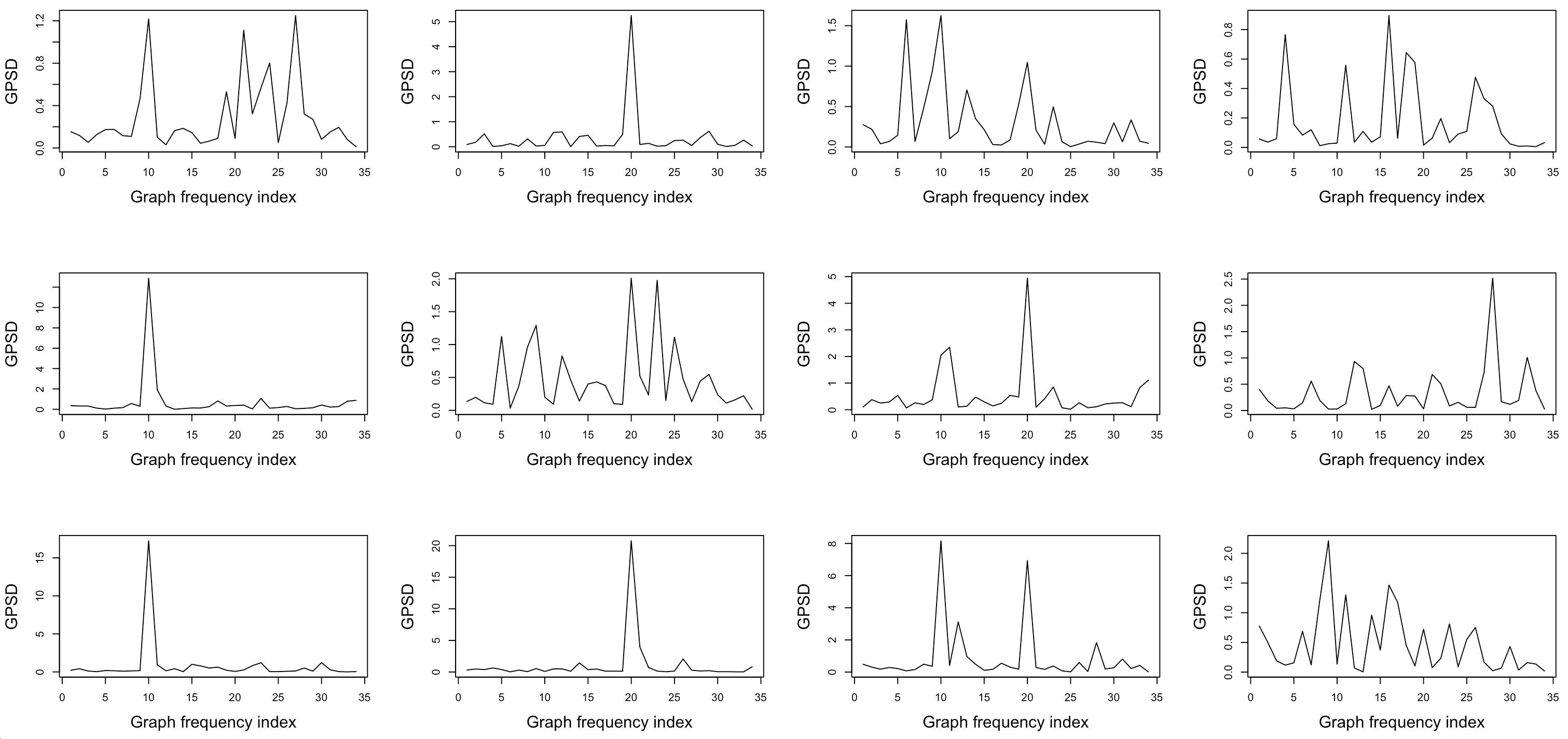}
        \vspace{-3mm}
    \caption{(From top to bottom, left to right) windowed average graph periodograms of signal $X_i$ ($i=1,\ldots,12$) generated on the Karate club network.}
    \label{fig:karate_gpsd}
\end{figure}

For the Karate club network, we generate a 12-dimensional graph signal. Let $v_\ell^{\text{KN}}$ and $\lambda_\ell^{\text{KN}}$ denote the $\ell$th eigenvector and eigenvalue of the graph Laplacian, respectively. The signal $X_i$ for dimension indices $1 \le i \le 12$ is defined as
\begin{equation*}
    \begin{aligned}
        X_i &= c_{i,1} v_{10}^{\text{KN}} + \delta_i, \quad i=1, 2, 3, \\
        X_i &= c_{i,2} v_{20}^{\text{KN}} + \delta_i, \quad i=4, 5, 6, \\
        X_i &= c_{i,1} v_{10}^{\text{KN}} + c_{i,2} v_{20}^{\text{KN}} + \delta_i, \quad i=7, 8, 9, \\
        X_i &= \delta_i, \quad i=10, 11, 12,
    \end{aligned}
\end{equation*}
where $\delta_i \sim N(0, 0.5^2)$ denote Gaussian noises. The amplitudes are given by $c_{1,1} = 1,~c_{2,1} = 2.5,~c_{3,1} = 3.5$, $c_{4,2} = 2,~c_{5,2} = 1.7,~c_{6,2} = 3.2$, $c_{7,1} = 2.1,~c_{7,2} = 0.9,~c_{8,1} = 1.4,~c_{8,2} = 2,~c_{9,1} = 2.5$, and $c_{9,2} = 2.2$. The windowed average graph periodograms for each $X_i$ are shown in \Cref{fig:karate_gpsd}. 

\begin{figure}[!t]
    \centering
        \includegraphics[width=0.95\textwidth]{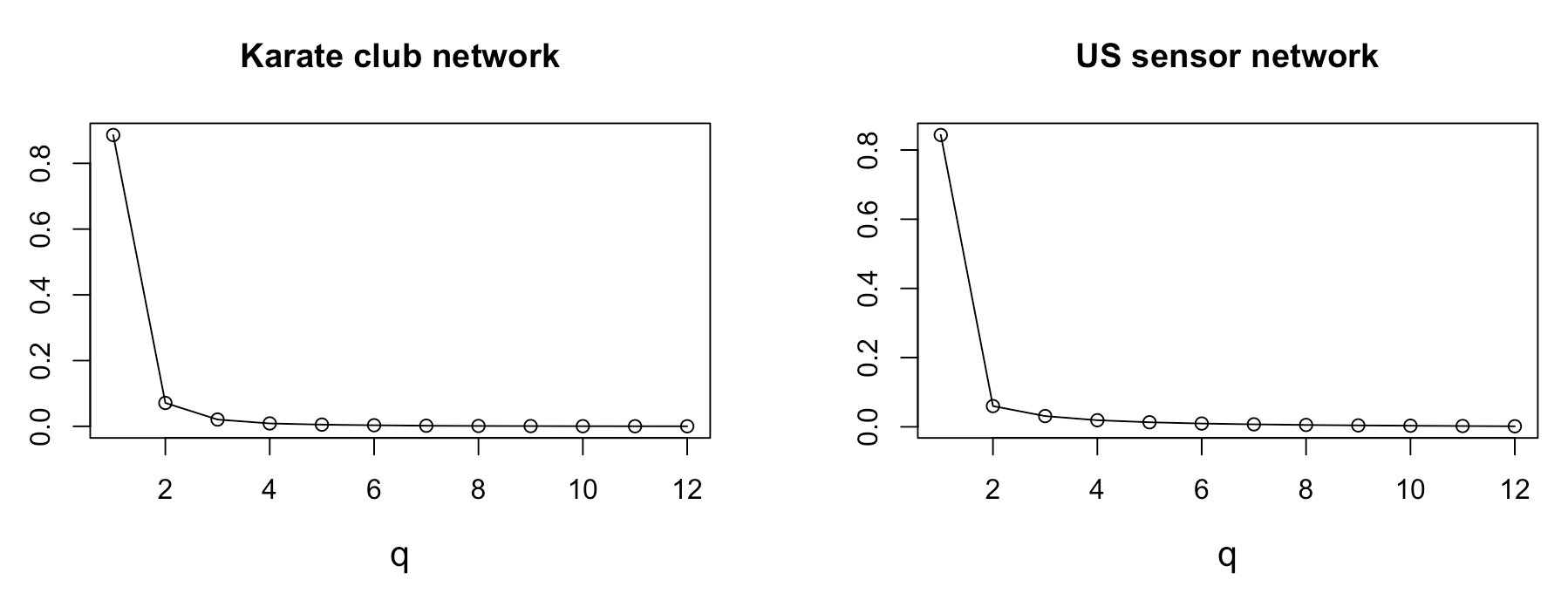}
        \vspace{-3mm}
    \caption{Scree plots for Karate club network (left) and US sensor network (right).}
    \label{fig:simul_screeplot}
\end{figure}

The scree plot in the left panel of \Cref{fig:simul_screeplot} is used to determine the number of reduced dimensions $q$ in the proposed gFreqPCA. The first and second principal components reduce the reconstruction error by 88.6\% and 7.1\%, respectively; therefore, the proportion of the cumulative reduction in the reconstruction error attributed to the two principal components exceeds 95\%, so we set $q=2$. 

For visual inspection, a simulated realization of $X_i$ ($i=1,\ldots,12$) on the Karate club network is shown in Figure C.1 of the supplementary material, and the corresponding 12 error signals (residuals), $X_i - \hat{X}_i$, obtained by two principal component graph signals, are displayed in Figure C.2 of the supplementary material. As one can see, the residuals visually become much smaller than the values of the original signal, indicating that the reconstruction is successful.

\begin{figure}[!t]
    \centering
        \includegraphics[width=0.95\textwidth]{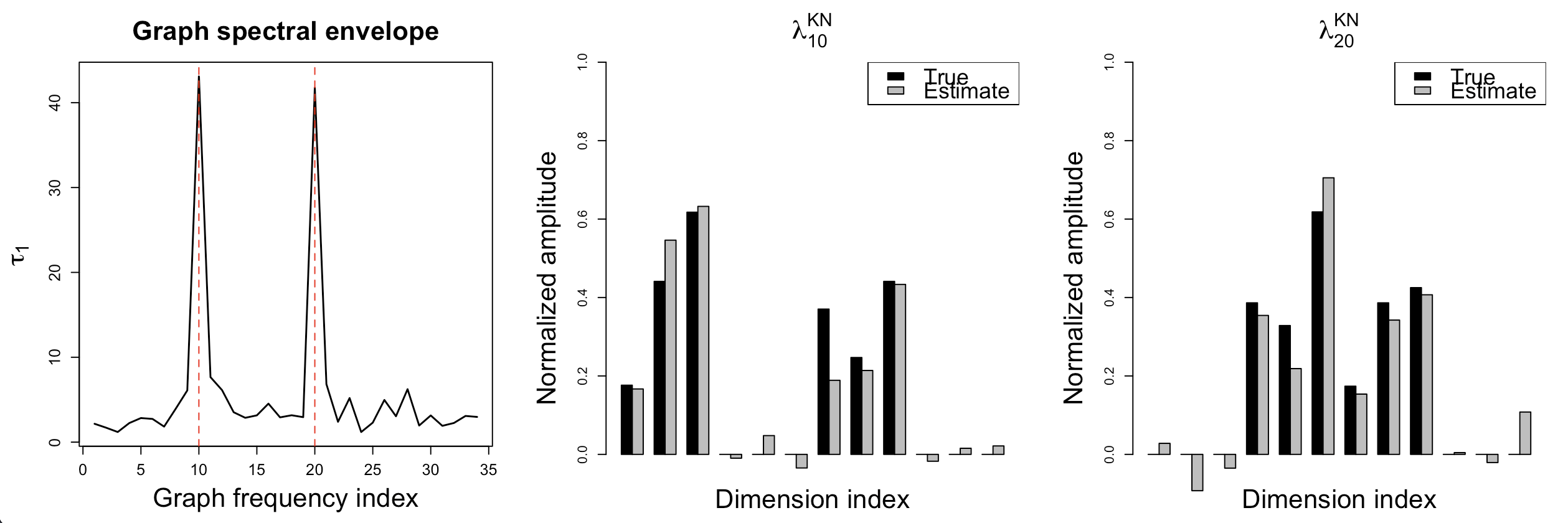}
        \vspace{-3mm}
    \caption{Results of gFreqPCA for Karate club network: Graph spectral envelope (left), and the optimal graph frequency scalings at $\lambda_{10}^{\text{KN}}$ (middle) and $\lambda_{20}^{\text{KN}}$ (right).}
    \label{fig:karate_amplitude}
\end{figure}

The resulting graph spectral envelope and the optimal graph frequency scalings are shown in \Cref{fig:karate_amplitude}. The graph spectral envelope accurately identifies two graph frequencies, $\lambda_{10}^{\text{KN}}$ and $\lambda_{20}^{\text{KN}}$, that represent the peaks. These frequencies correspond to the eigenvectors $v_{10}^{\text{KN}}$ and $v_{20}^{\text{KN}}$, which were used to generate the signal. Furthermore, the optimal graph frequency scalings at $\lambda_{10}^{\text{KN}}$ and $\lambda_{20}^{\text{KN}}$ are shown in \Cref{fig:karate_amplitude}. They successfully estimate the normalized amplitudes derived from $c_{1,1}$ to $c_{9,2}$, demonstrating the superiority of the proposed gFreqPCA described in \Cref{sec:gfreqPCA}. 

\begin{figure}[!t]
    \centering
        \includegraphics[width=0.95\textwidth]{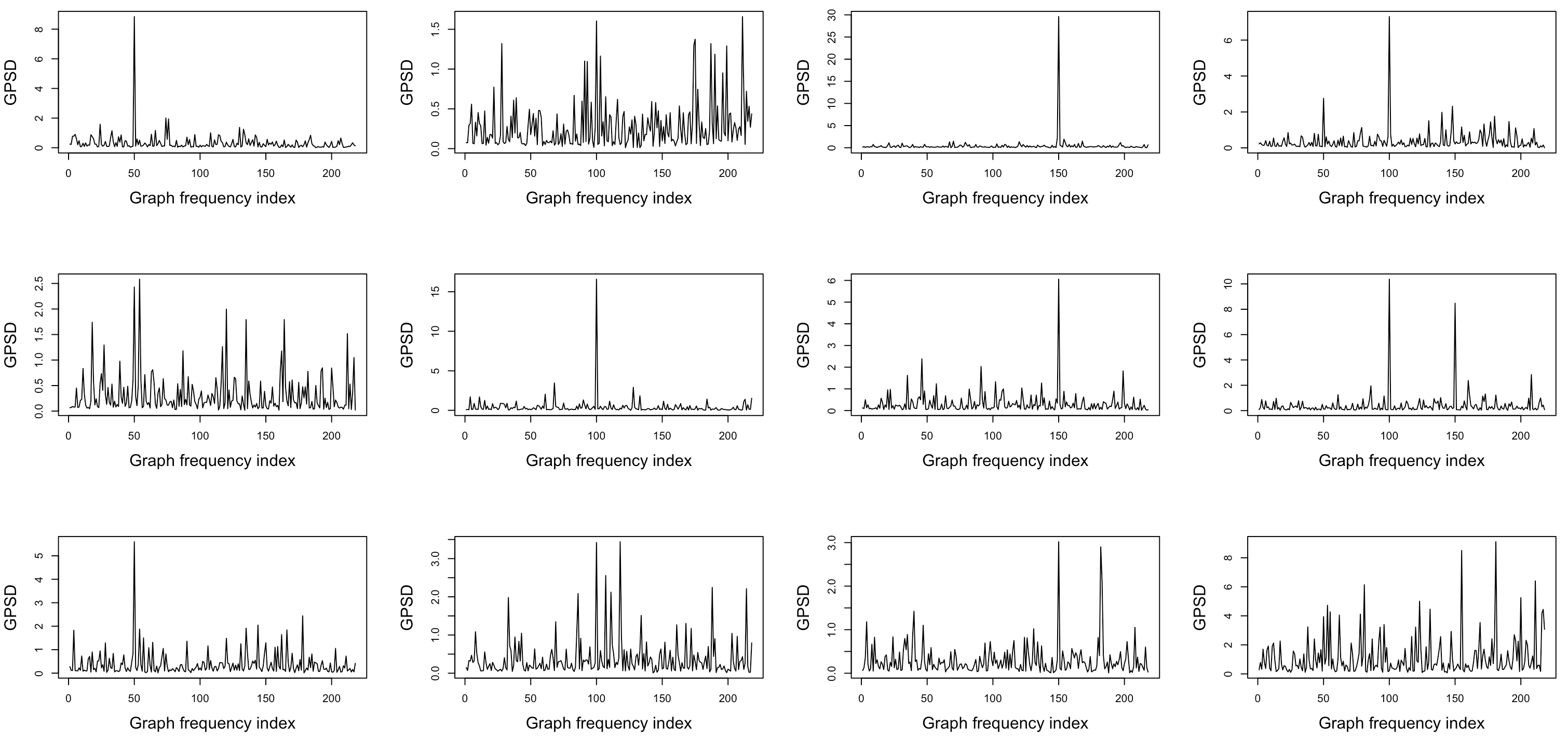}
        \vspace{-3mm}
    \caption{(From top to bottom, left to right) windowed average graph periodograms of of signal $X_i$ ($i=1,\ldots,12$) generated on the US sensor network.}
    \label{fig:ustemp_gpsd}
\end{figure}

Similarly, for the US sensor network, we generate a 12-dimensional graph signal. Let $v_\ell^{\text{US}}$ and $\lambda_\ell^{\text{US}}$ denote the $\ell$th eigenvector and eigenvalue of the graph Laplacian, respectively. The signal $X_i$ for dimension indices $1 \le i \le 12$ is defined as
\begin{equation*}
    \begin{aligned}
        X_i &= c_{i,1} v_{50}^{\text{US}} + \delta_i, \quad i=1, 2, 3, \\
        X_i &= c_{i,2} v_{100}^{\text{US}} + \delta_i, \quad i=4, 5, 6, \\
        X_i &= c_{i,3} v_{150}^{\text{US}} + \delta_i, \quad i=7, 8, 9, \\
        X_{i} &= c_{i,1} v_{50}^{\text{US}} + c_{i,2} v_{100}^{\text{US}} + \delta_i, \quad i=10, \\
        X_{i} &= c_{i,2} v_{100}^{\text{US}} + c_{i,3} v_{150}^{\text{US}} + \delta_i, \quad i=11, \\
        X_{i} &= 2\delta_i, \quad i=12,
    \end{aligned}
\end{equation*}
where $\delta_i \sim N(0, 0.5^2)$ denote Gaussian noises. The amplitudes are given by $c_{1,1} = 3,~c_{2,1} = 1.5,~c_{3,1} = 2$, $c_{4,2} = 2,~c_{5,2} = 4,~c_{6,2} = 3$, $c_{7,3} = 5,~c_{8,3} = 2,~c_{9,3} = 1.5$, $c_{10,1} = 2,~c_{10,2} = 4$, $c_{11,2} = 3$, and $c_{11,3} = 2.5$. The windowed average graph periodograms for each $X_i$ are shown in \Cref{fig:ustemp_gpsd}.

From the scree plot in the right panel of \Cref{fig:simul_screeplot}, the first four principal components reduce the reconstruction error by 84.3\%, 6.0\%, 3.1\%, and 1.9\%, respectively. Thus, the proportion of the cumulative reduction in the reconstruction error attributed to these four principal components exceeds 95\%, so we set $q=4$. 

\begin{figure}[!t]
    \centering
        \includegraphics[width=0.95\textwidth]{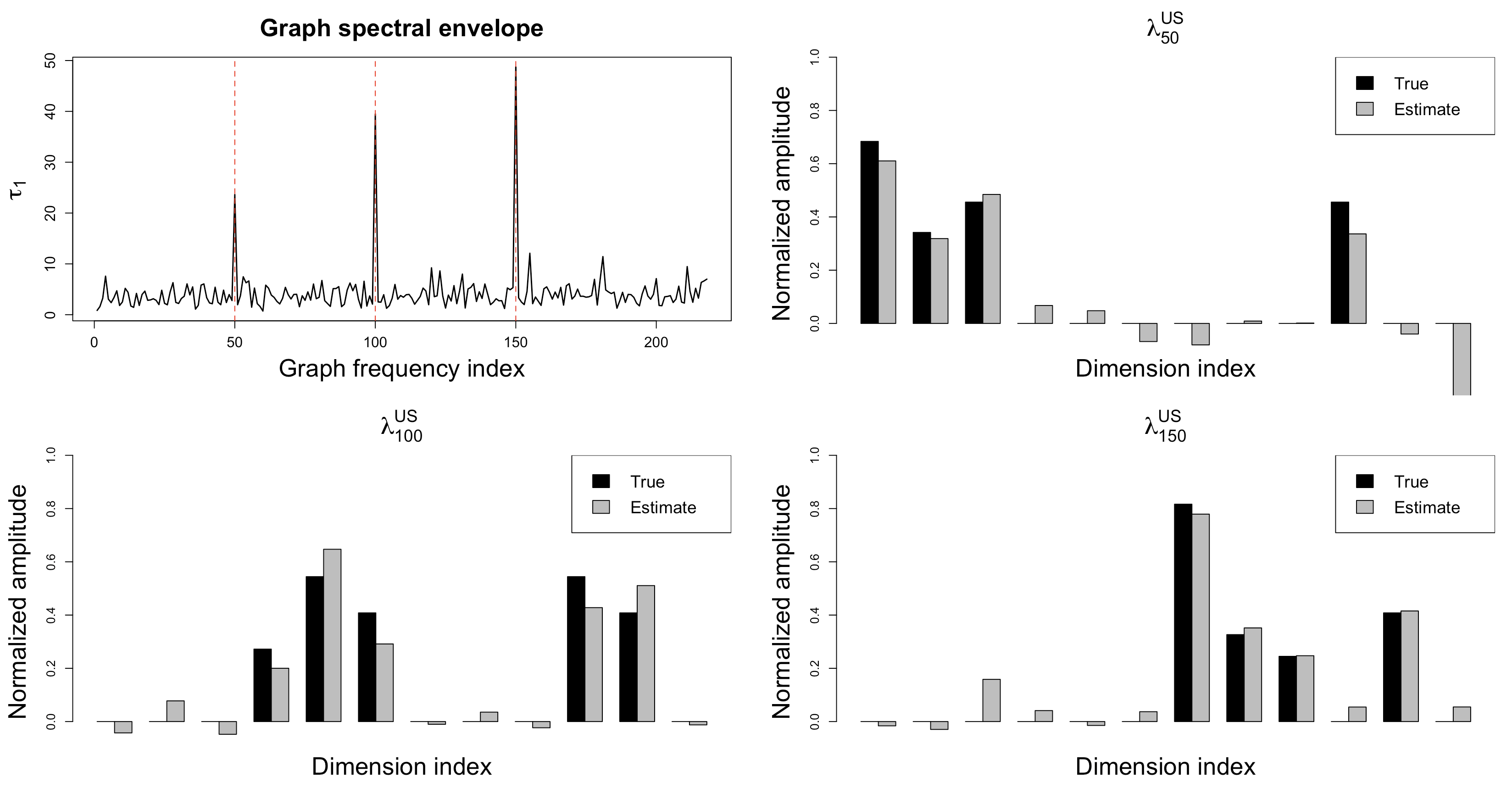}
        \vspace{-3mm}
    \caption{Results of PCA in the graph frequency domain for Karate club network: Graph spectral envelope (top left), and the optimal graph frequency scalings at $\lambda_{50}^{\text{US}}$ (top right),  $\lambda_{100}^{\text{US}}$ (bottom left), and $\lambda_{150}^{\text{US}}$ (bottom right).}
    \label{fig:ustemp_amplitude}
\end{figure}

\Cref{fig:ustemp_amplitude} shows the resulting graph spectral envelope and the optimal graph frequency scalings. As one can see, the graph spectral envelope successfully identifies three graph frequencies, $\lambda_{50}^{\text{US}}$, $\lambda_{100}^{\text{US}}$, and $\lambda_{150}^{\text{US}}$, where distinct peaks are observed. These frequencies correspond to the eigenvectors $v_{50}^{\text{US}}$, $v_{100}^{\text{US}}$, and $v_{150}^{\text{US}}$ used to generate the signal. Furthermore, the optimal graph frequency scalings at $\lambda_{50}^{\text{US}}$, $\lambda_{100}^{\text{US}}$, and $\lambda_{150}^{\text{US}}$ accurately  estimate the normalized amplitudes derived from $c_{1,1}$ to $c_{11,3}$, highlighting the effectiveness of the proposed gFreqPCA. 

\subsection{Real Data Analysis} \label{sec:realdataanalysis}
\subsubsection{Seoul Metropolitan Subway Data} \label{sec:metrodata}
For real-world data analysis, we use Seoul Metropolitan subway data, which contains hourly information about the number of passengers getting on and off at different stations of the Seoul Metro in South Korea. The original data are publicly available on the Seoul Subway website (\url{http://www.seoulmetro.co.kr}). Subway data inherently forms a graph, with stations representing the vertices and railroads representing the edges. For this analysis, we focus on 243 stations with data from January 2, 2021, to November 30, 2021. In particular, we analyze data from Line 2, one of the eight lines in the dataset. Line 2 is unique in that it is the only circular line in the Seoul Metro system that runs through major areas of Seoul, including business districts, commercial centers, university towns, and residential places. This characteristic of Line 2 makes it particularly interesting for our analysis. Figure 7 shows Line 2 on a map of Seoul, with orange tones representing the average 2021 resident population: darker tones indicate more population, and lighter tones represent less population. The blue curve in the middle of Seoul marks the Han River.

\begin{figure}[!t]
    \centering
        \includegraphics[width=0.8\textwidth]{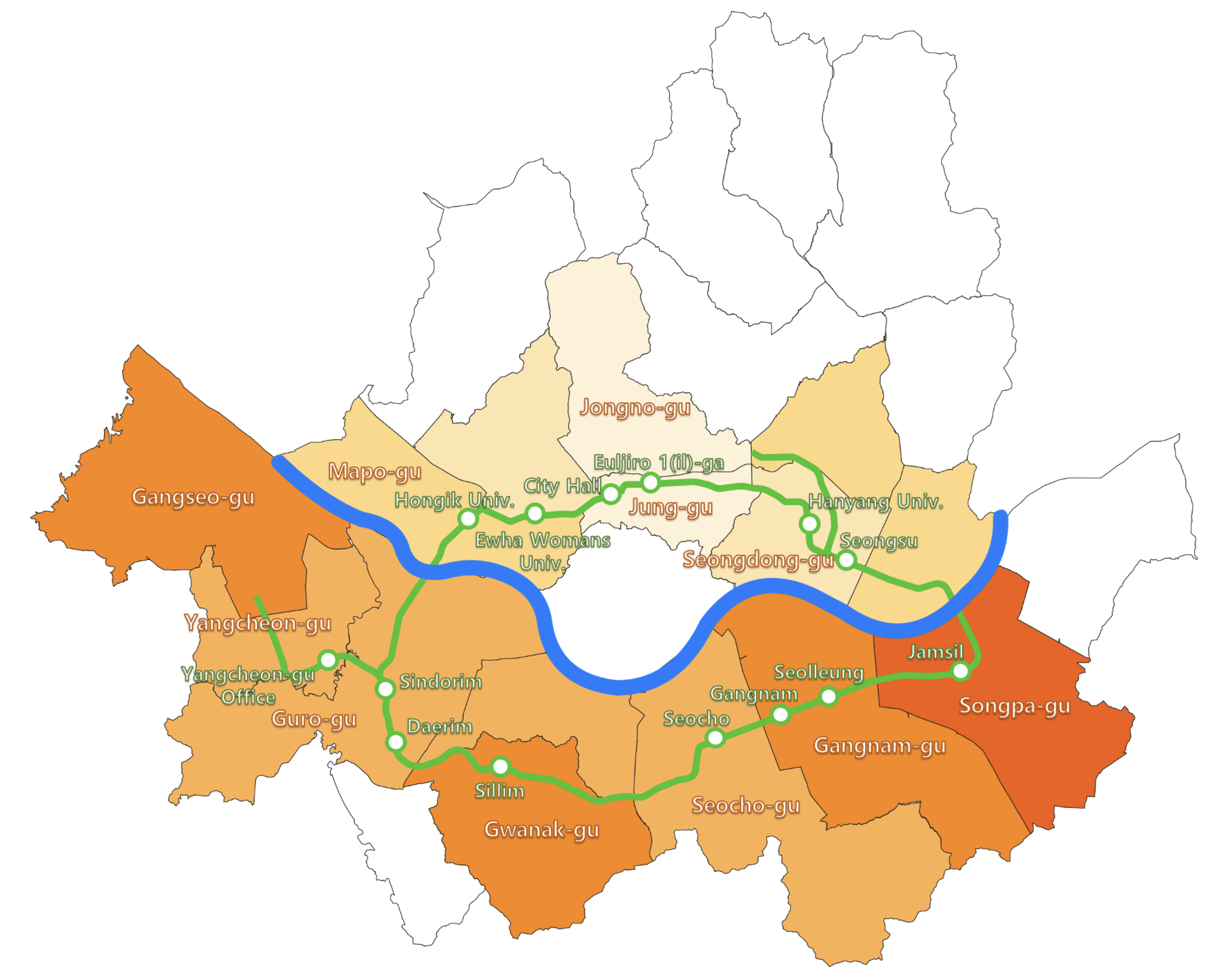}
        \vspace{-3mm}
    \caption{Map of Seoul showing Line 2 of the Seoul Metro system (green), with major stations and districts it passes through, where the blue curve marks the Han River. The resident registration population is represented in orange tones.}
    \label{fig:seoulmetro_hanriver}
\end{figure}

To construct a weighted graph, we assign edge weights based on the distance between stations. Specifically, the edge weight between station $i$ and station $j$ is determined by $w_{ij} = \exp(-d^2(i,j) / ave^2)$, where $d(i,j)$ represents the distance between station $i$ and station $j$ calculated by summing the lengths of the railroads connecting the two stations, and $ave$ denotes the average distance between all stations.

We divided the day into several time intervals: early morning (before 6 a.m.), rush (morning) hour (6 a.m. - 10 a.m.), work hours (10 a.m. - 4 p.m.), rush (afternoon) time (4 p.m. - 8 p.m.), and night time (after 8 p.m.), denoted by $T_1, \ldots, T_5$, respectively. For each station, we averaged the data within these intervals, resulting in five averaged values representing the number of passengers boarding and alighting. These values form a 5-dimensional graph signal at each station. Here, we consider two separate graph signals: one representing the number of passengers boarding and the other the number of passengers alighting. This allows us to analyze the different behaviors of passengers at different stations based on their activity type. We applied a log transformation $x \mapsto \log(1+x)$ to each signal value at each station and each time interval and then centered the graph signals in each dimension to ensure a mean of zero across the stations for convenience.

\subsubsection{Understanding Seoul's Urban Structure for Analysis} \label{sec:understandingseoul}
\begin{figure}[!t]
    \centering
        \includegraphics[width=0.9\textwidth]{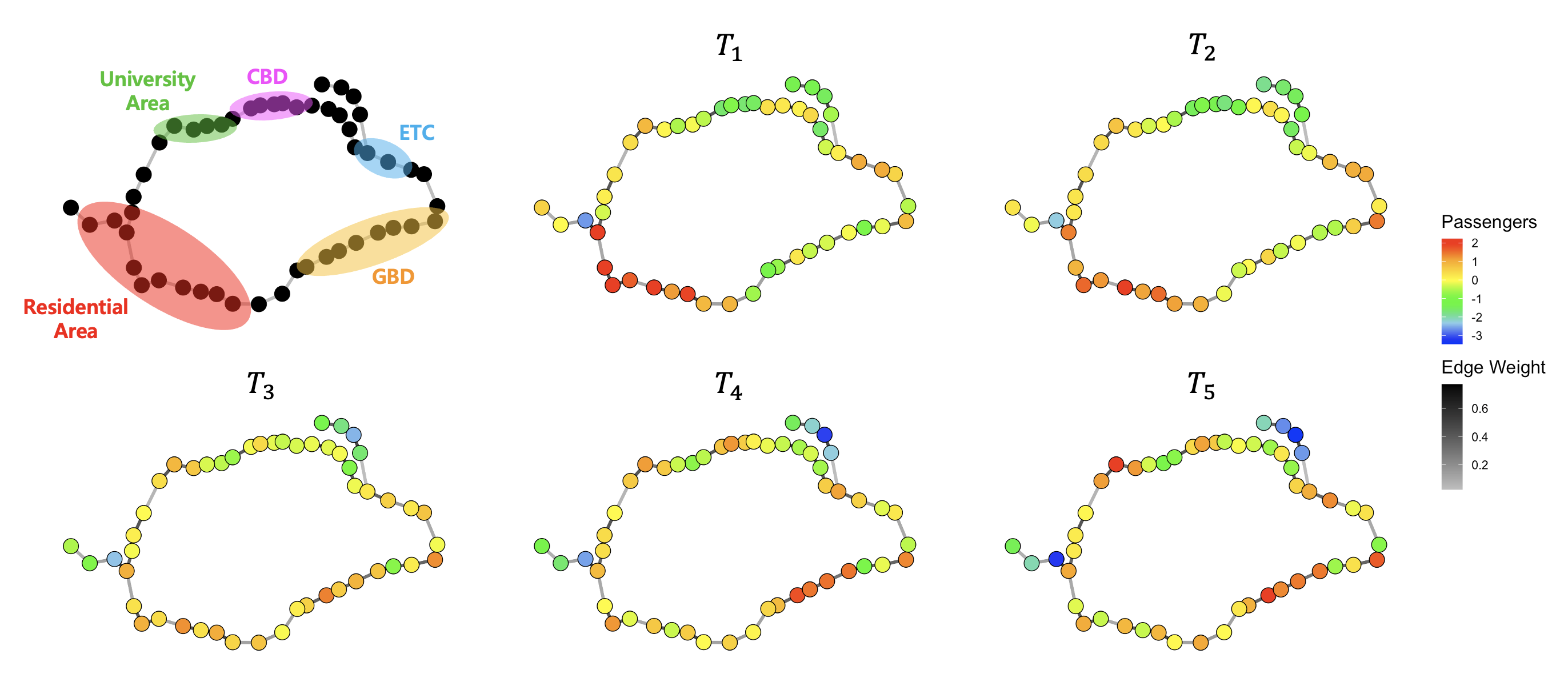}
        \vspace{-3mm}
    \caption{Log-transformed and centered data of boarding passengers: Graph of Line 2 with regional characteristics (top left). Signals for each time interval $T_1, \ldots, T_5$ (remaining panels).}
    \label{fig:seoulmetro_data_getin}
\end{figure}

\begin{figure}[!t]
    \centering
        \includegraphics[width=0.9\textwidth]{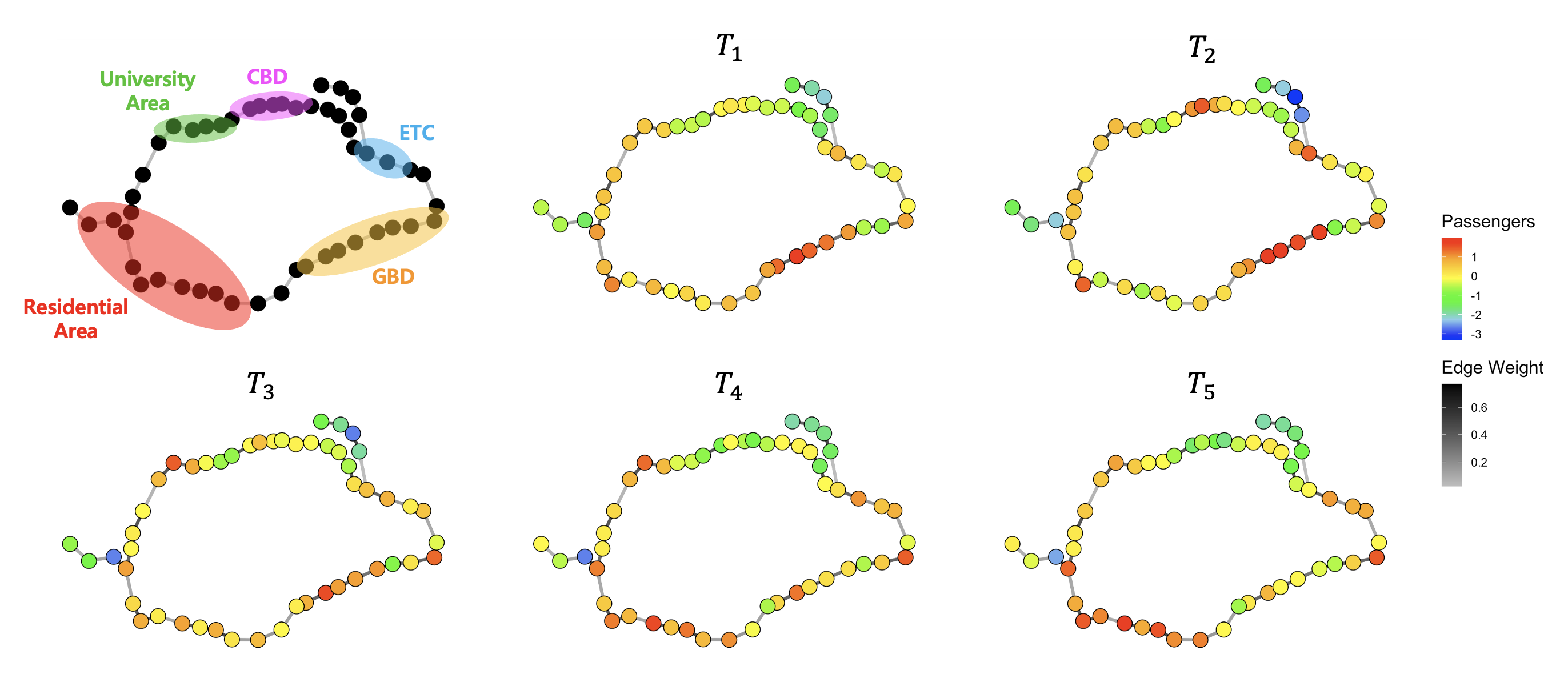}
        \vspace{-3mm}
    \caption{Log-transformed and centered data of alighting passengers: Graph of Line 2 with regional characteristics (top left). Signals for each time interval $T_1, \ldots, T_5$ (remaining panels).}
    \label{fig:seoulmetro_data_getout}
\end{figure}

The two resulting 5-dimensional graph signals are shown in Figures \ref{fig:seoulmetro_data_getin} and \ref{fig:seoulmetro_data_getout}. To understand the patterns of boarding and alighting passengers, we first need to look at the urban structure of Seoul. In terms of residential characteristics, more people live south of the Han River, as shown in Figure 7. This is mainly due to ongoing redevelopment and reconstruction projects, as well as the perception that the southern part of the city has better education and work infrastructure than the northern part. As a result, there is an influx of people to the southern part of Seoul. 

Among the districts south of the Han River, Gangnam-gu, Seocho-gu, and Songpa-gu, located in the east, rank in the top five in average wage per capita. Residents of these high-income neighborhoods are more likely to commute by car than by subway, unlike residents of Gangseo-gu and Gwanak-gu, which are among the top five districts by registered population. Given this and the population density, there is likely to be high demand for subway service in the southwestern part of Seoul, colored red and labeled as `Residential Area' in the top left panel of \Cref{fig:seoulmetro_data_getin}. 

Regarding the working environment, Seoul has three main business districts: Gangnam Business District (GBD), Central Business District (CBD), and Yeouido Business District (YBD). Of these, Line 2 runs through the GBD and CBD. The GBD covers the neighborhoods of Gangnam-gu, Seocho-gu, and Songpa-gu. It is a hub for small and medium-sized business-to-business (B2B) firms and thriving startup ecosystems, mainly IT companies. The CBD covers the business districts around Jongno-gu and Jung-gu. It is the oldest business district centered around City Hall. It hosts many government and public institutions, such as the Bank of Korea, large corporate headquarters, and foreign companies. In addition, there is a university area in the west of the CBD, including Hongik University, Ewha Womans University, Yonsei University, and Sogang University, which are marked in the upper left panel of \Cref{fig:seoulmetro_data_getin}. Moreover, the area around Seongsu-dong, labeled ETC in the top left panel of \Cref{fig:seoulmetro_data_getin}, is an emerging business district known for its fashion and K-POP culture, attracting a younger generation. This area is often called ``Korea's Brooklyn" and has begun attracting many startup offices and tech companies.

The urban structure of Seoul described above clearly explains the patterns observed in Figures \ref{fig:seoulmetro_data_getin} and \ref{fig:seoulmetro_data_getout}. We can see that in the early morning and during rush (morning) hours, many passengers board at stations in the southwestern residential areas and alight at stations in the northern and eastern regions, such as the CBD, GBD, university area, and Seongsu-dong, to get to work. Conversely, during rush (afternoon) hours and night time, many passengers board at stations in the northern and eastern areas and alight at stations in the southwestern residential areas to return home.

\subsubsection{Results of Analysis} \label{sec:realdataresults}
We apply the proposed gFreqPCA to the Seoul Metro Line 2 data. We set the number of windows $M=50$. For both graph signals representing boarding and alighting passengers, the first principal component graph signal reduces the reconstruction error by 99.6\%, so we set $q=1$.

\begin{figure}[!ht]
    \centering
        \includegraphics[width=0.9\textwidth]{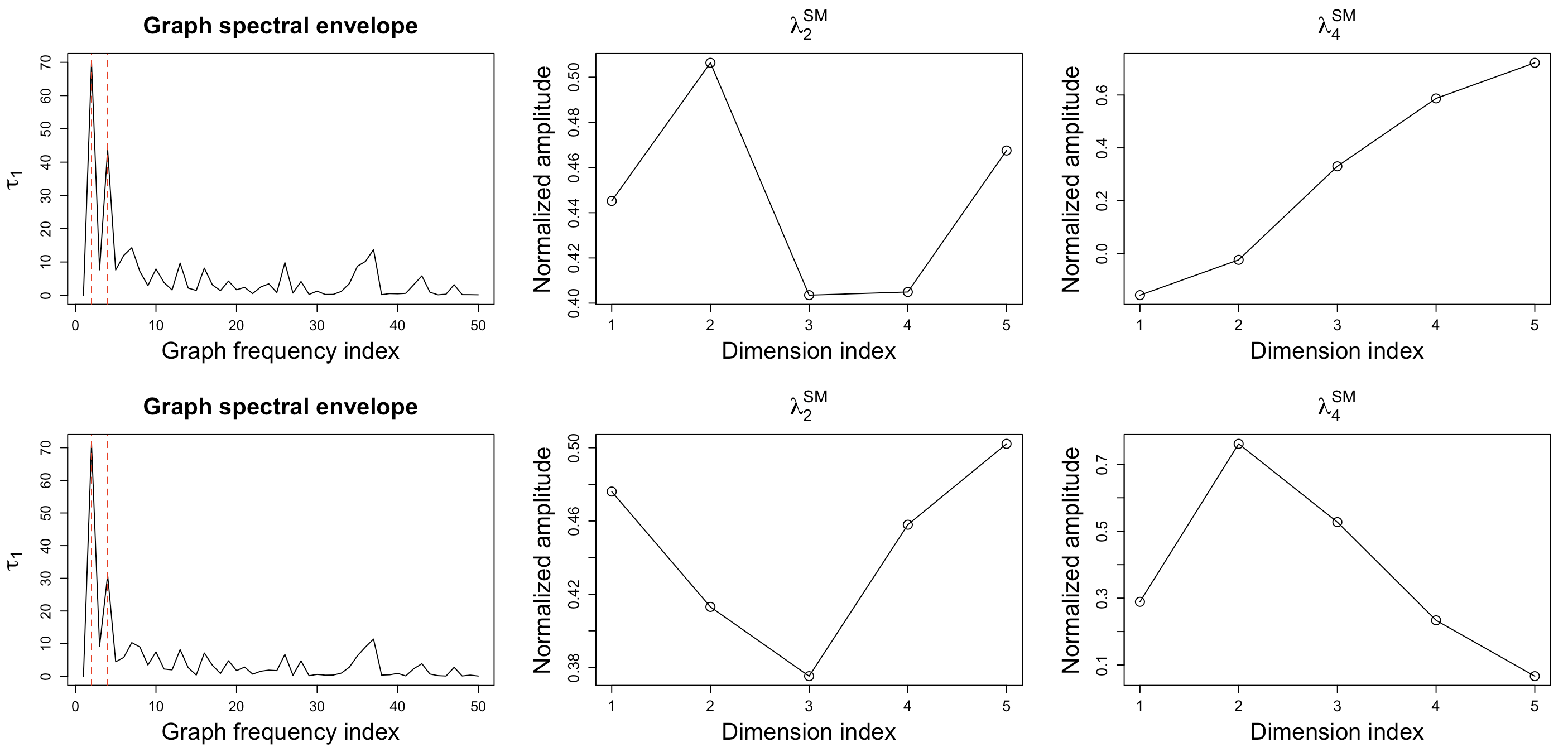}
        \vspace{-3mm}
    \caption{Results of PCA in the graph frequency domain for boarding passengers data (top row) and alighting passengers data (bottom row): Graph spectral envelopes (left), and the optimal graph frequency scalings at $\lambda_2^{\text{SM}}$ (middle) and $\lambda_4^{\text{SM}}$ (right).}
    \label{fig:metro_amplitude}
\end{figure}

\begin{figure}[!ht]
    \centering
        \includegraphics[width=0.9\textwidth]{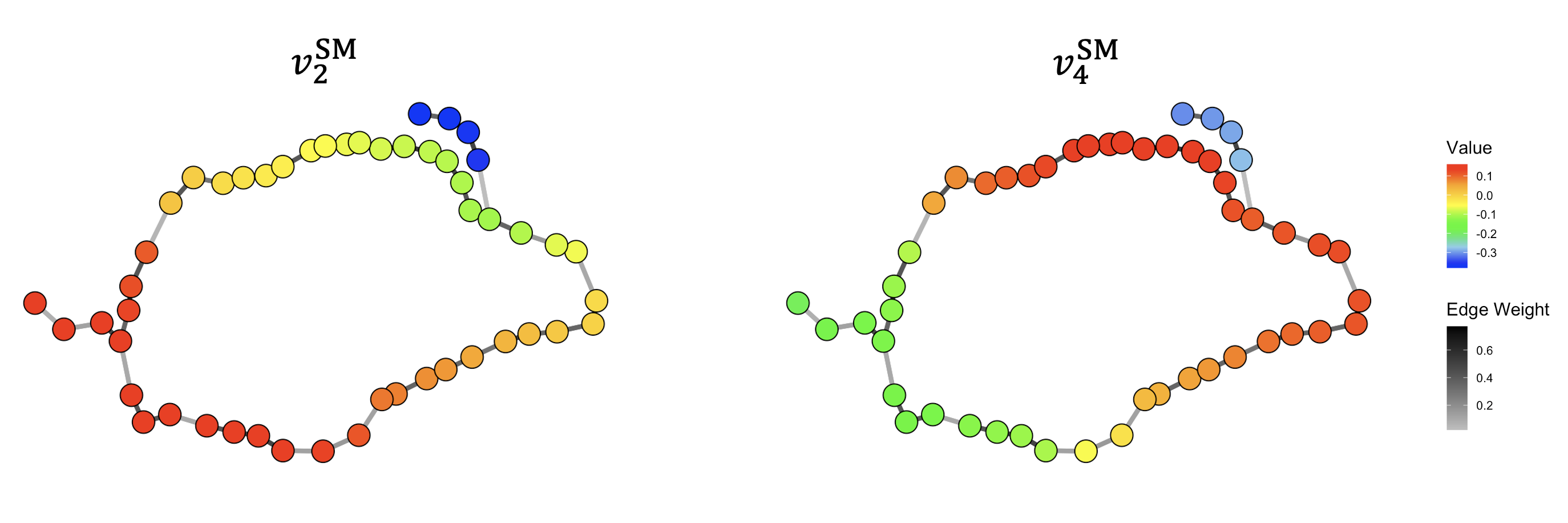}
        \vspace{-3mm}
    \caption{Signals of the eigenvectors $v_{2}^{\text{SM}}$ and $v_{4}^{\text{SM}}$.}
    \label{fig:seoulmetro_evectors}
\end{figure}

The graph spectral envelopes for both datasets, shown in the left column of \Cref{fig:metro_amplitude}, exhibit two prominent peaks at $\lambda_2^{\text{SM}}$ and $\lambda_4^{\text{SM}}$, where $\lambda_i^{\text{SM}}$ represents the $i$th graph frequency of the Seoul Metropolitan subway graph, corresponding to the $i$th eigenvalue of the graph Laplacian. This indicates that the significant graph frequencies $\lambda_2^{\text{SM}}$ and $\lambda_4^{\text{SM}}$ exist in the 5-dimensional graph signal for both boarding and alighting passenger data. The corresponding eigenvectors $v_{2}^{\text{SM}}$ and $v_{4}^{\text{SM}}$ are shown in \Cref{fig:seoulmetro_evectors}. The signal values of $v_{2}^{\text{SM}}$ at southwest stations are higher than those at northeast stations, suggesting that $v_{2}^{\text{SM}}$ is associated with residential population density. Conversely, the signal values of $v_{4}^{\text{SM}}$ at northeast stations, excluding those on the branch line marked in blue, are higher than those at southwest stations. This implies that $v_{4}^{\text{SM}}$ relates to working areas, including office districts and university areas.

We estimate the normalized amplitudes corresponding to  $v_{2}^{\text{SM}}$ and $v_{4}^{\text{SM}}$ by calculating the optimal graph frequency scalings at $\lambda_{2}^{\text{SM}}$ and $\lambda_{4}^{\text{SM}}$, respectively. The results are shown in \Cref{fig:metro_amplitude}. Denote the optimal graph frequency scaling at $\lambda_i^{\text{SM}}$ by $u_1^{\text{SM}}(\lambda_i^{\text{SM}})$. For the boarding passengers signal, $u_1^{\text{SM}}(\lambda_2^{\text{SM}})$ shows a relatively high value at $T_2$ (dimension index 2) and a relatively low value at $T_4$ (dimension index 4) compared to other time intervals, whereas $u_1^{\text{SM}}(\lambda_4^{\text{SM}})$ exhibits a relatively low value at $T_2$ and a relatively high value at $T_4$. This indicates that the graph signal at $T_2$ has a higher amplitude for $v_{2}^{\text{SM}}$ than for $v_{4}^{\text{SM}}$, while at $T_4$, the graph signal has a higher amplitude for $v_{4}^{\text{SM}}$ than for $v_{2}^{\text{SM}}$. These findings align with the expectations inferred from \Cref{sec:understandingseoul}. Conversely, for the alighting passengers signal, $u_1^{\text{SM}}(\lambda_2^{\text{SM}})$ shows a relatively lower value than $u_1^{\text{SM}}(\lambda_4^{\text{SM}})$ at $T_2$, while at $T_4$, $u_1^{\text{SM}}(\lambda_2^{\text{SM}})$ shows a relatively higher value than $u_1^{\text{SM}}(\lambda_4^{\text{SM}})$. This suggests that at $T_2$, the graph signal  has a higher amplitude for $v_{4}^{\text{SM}}$ than for $v_{2}^{\text{SM}}$, whereas at $T_4$, the graph signal has a higher amplitude for $v_{2}^{\text{SM}}$ than for $v_{4}^{\text{SM}}$.



\section{Concluding Remarks} \label{sec:conclude}
In this paper, we proposed gFreqPCA, a novel principal component analysis method in the graph frequency domain, for reducing the dimensionality of multivariate data on graphs. Unlike many existing dimensionality reduction methods on graphs, the proposed gFreqPCA provides an explicit form of the dimension-reduced signal and the reconstructed signal without requiring an optimization process. It also provides the graph spectral envelope and the optimal graph frequency scaling. The graph spectral envelope enables the identification of common graph frequencies within a multivariate graph signal, and the optimal graph frequency scaling allows for the estimation of the normalized amplitude of the eigenvectors that form the bases of the signal for each dimension. We validated the effectiveness of the proposed method through various simulations and applied it to Seoul Metropolitan Subway data to analyze passenger boarding and alighting patterns.

Although the proposed method is a novel dimensionality reduction method for multivariate graph signals, it still needs some improvements. Since it assumes the stationarity of multivariate graph signals, a dimensionality reduction method for non-stationary graph signals needs to be developed. Further improvements to the estimation method of GCSD are also needed. For example, we need to study the optimal window design for the windowed average graph cross-periodogram. We leave these topics for future research.

\section*{Acknowledgments}
This research was supported by the National Research Foundation of Korea (NRF) funded by the Korea government (2021R1A2C1091357).

\clearpage
\def\thesection{\Alph{section}}
\counterwithin{equation}{section}
	\begin{appendices}
 \section{Dimension Reduction Methods on Graphs} \label{sec:relatedworks}
Several studies have investigated low-dimensional representations of a multivariate graph signal on $\mc{G}$, denoted as an $n \times p$ matrix $X=(X_1 \mid \cdots \mid X_p)$, where each $X_i$ ($1 \le i \le p$) is an $n$-dimensional vector.
\citet{Jiang2013graph} proposed a graph Laplacian PCA (gLPCA) formulated as 
\begin{equation*}
    \underset{U,Q}{\min} \; \lVert X^\top - UQ^\top \rVert_F^2 + \alpha \operatorname{tr}[Q^\top L Q] \quad \text{s.t.} \quad  Q^\top Q = I,
\end{equation*}
where $\lVert \cdot \rVert_F$ denotes the Frobenius norm,  $\alpha \ge 0$ is a parameter that balances the contributions of the two terms, $L$ is the graph Laplacian matrix, and $U$ and $Q$ are $p \times q$ and $n \times q$ matrices, respectively. The former term is equivalent to the error minimized in classical PCA, while the latter term corresponds to the error minimized in Laplacian embedding that aims to maximize smoothness with respect to the underlying graph $\mc{G}$  \citep{Belkin2001laplacian, Belkin2003laplacian}. Each row of $Q$ represents the dimension-reduced vector corresponding to each vertex, serving as its principal component. The solution of gLPCA is given by $\hat{Q}$, which is a matrix of the $q$ eigenvectors corresponding to the first $q$ smallest eigenvalues of the matrix $(-XX^\top + \alpha L)$, and $\hat{U}=X^\top\hat{Q}$.

A robust version of gLPCA (RgLPCA) was also proposed by \citet{Jiang2013graph} using the $L_{2,1}$ norm, defined as $\lVert A \rVert_{2,1} = \sum_{j=1}^n \sqrt{\sum_{i=1}^p A_{ij}^2}$ for a $p \times n$ matrix $A$. The optimization problem is formulated as 
\begin{equation*}
    \underset{U,Q}{\min} \; \lVert X^\top - UQ^\top \rVert_{2,1}^2 + \alpha \operatorname{tr}[Q^\top L Q] \quad \text{s.t.} \quad  Q^\top Q = I.
\end{equation*}
The optimizing solution can be evaluated using the Augmented Lagrange Multiplier (ALM) method.

\citet{Cai2010graph} proposed the graph regularized nonnegative matrix factorization (GNMF), defined by the following optimization problem,
\begin{equation*}
    \underset{U \ge 0, Q \ge 0}{\min} \; \lVert X^\top - UQ^\top \rVert_{F}^2 + \alpha \operatorname{tr}[Q^\top L Q],
\end{equation*}
where $\alpha$ controls the smoothness of the dimension-reduced vectors. The former term in this formulation comes from the nonnegative matrix factorization (NMF) technique, which requires $U$ and $Q$ to be nonnegative matrices, ensuring that their elements are nonnegative for physical interpretation. The latter term serves as a regularization term that accounts for the local invariance on the graph. The optimizing solution is obtained through the iterative algorithm described in \citet{Cai2010graph}. 

\citet{Zhang2012low} proposed the manifold regularized matrix factorization (MMF) formulated as
\begin{equation*}
    \underset{U,Q}{\min} \; \lVert X^\top - UQ^\top \rVert_F^2 + \alpha \operatorname{tr}[Q^\top L Q] \quad \text{s.t.} \quad  U^\top U = I,
\end{equation*}
where the orthogonality constraint is imposed on $U$ rather than $Q$. It is guaranteed to have a globally optimal solution, which can be found using an iterative algorithm.

\citet{Shahid2015robust} proposed a robust PCA on graphs (RPCAG), defined as
\begin{equation*}
    \underset{\mc{L}, \mc{S}}{\min} \; \lVert \mc{L} \rVert_* + \alpha_1 \lVert \mc{S} \rVert_1 + \alpha_2 \operatorname{tr}[\mc{L} L \mc{L}^\top] \quad \text{s.t.} \quad  X^\top = \mc{L} + \mc{S},
\end{equation*}
where $\lVert \cdot \rVert_*$ denotes the nuclear norm, $\mc{L}$ and $\mc{S}$ represent the low-rank approximation and sparse errors, respectively, and parameters $\alpha_1$ and $\alpha_2$ control the degree of sparsity in $\mc{S}$ and the smoothness of $\mc{L}$ on the graph, respectively. \citet{Shahid2016fast} introduced a fast, robust PCA on graphs (FRPCAG) without using the expensive nuclear norm in RPCAG as
\begin{equation*}
    \underset{\mc{L}, \mc{S}}{\min} \; \lVert \mc{S} \rVert_1  + \alpha_1 \operatorname{tr}[\mc{L} L^{(1)} \mc{L}^\top] + \alpha_2 \operatorname{tr}[\mc{L}^\top L^{(2)} \mc{L}]\quad \text{s.t.} \quad  X^\top = \mc{L} + \mc{S},
\end{equation*}
where $L^{(1)}$ and $L^{(2)}$ are graph Laplacian matrices corresponding to the two types of graphs $\mc{G}^{(1)}$ and $\mc{G}^{(2)}$, constructed between the data samples (columns of $X^\top$) and between the features (rows of $X^\top$), respectively. 

\citet{Shen2017nonlinear} proposed a kernel PCA on graphs, formulated as 
\begin{equation*}
    \underset{Q}{\min} \;  \operatorname{tr}[Q^\top K_x^{-1} Q] + \alpha \operatorname{tr}[Q^\top L Q]\quad \text{s.t.} \quad  Q^\top Q = \Gamma_q,
\end{equation*}
where $\Gamma_q$ is a $q \times q$ diagonal matrix containing the $q$ largest eigenvalues of $(K_x^{-1}+\alpha L)$, and $K_x$ is defined as $K_x = ((I-W)(I-W)^\top)^\dag$, with $\dag$ representing the pseudo inverse.

The dimensionality reduction methods we have seen are implemented in the vertex domain. On the other hand, the proposed method described in this paper implements dimensionality reduction in the frequency domain of the graph. The proposed approach can provide additional insights that cannot be obtained from vertex domain dimension reduction methods.

    \section{Proofs} \label{appendix:proof}
    \subsection{Proof of Proposition 3.1} \label{pf:psd_Plambda}
    \begin{proof}
    It is clear that $P_X(\lambda_\ell)$ is Hermitian, as $p_{ij}^X(\lambda_\ell) = p_{ji}^{X^*}(\lambda_\ell)$ \citep{Kim2024cross}.
    For the cross-covariance matrix $\Sigma_{ij}^X$ between $X_i$ and $X_j$, the GCSD $p_{ij}^X(\lambda_\ell)$ is given by $p_{ij}^X(\lambda_\ell) = v_\ell^H \Sigma_{ij}^X v_\ell$, where $v_\ell$ is the $\ell$th column of $V$. Let $\tilde{V}_\ell$ be a $\tilde{N} \times p$ matrix, where $\tilde{N}:=n\times p$, defined as
    \begin{equation*}
        \tilde{V}_\ell = \left(\begin{array}{cccc}
        v_\ell& \mathbf{0} & \cdots & \mathbf{0} \\
        \mathbf{0} & v_\ell & \cdots & \mathbf{0} \\
        \vdots & \vdots & \ddots & \vdots \\
        \mathbf{0} & \mathbf{0} & \cdots & v_\ell
        \end{array}\right),
            \end{equation*}
            where $\mathbf{0}$ represents an $n \times 1$ zero vector.
            Also, let $\tilde{\Sigma}^X$ be a $\tilde{N} \times \tilde{N}$ block matrix, where the $(i,j)$th block element is $\Sigma_{ij}^X$. Given this, the matrix $P_X(\lambda_\ell)$ can be expressed as $P_X(\lambda_\ell) = \tilde{V}_\ell^H \tilde{\Sigma}^X \tilde{V}_\ell$. For a vector $y \in \mathbb{C}^p$, let $\tilde{V}_\ell y$ be denoted by $\tilde{y}=(\tilde{y}_1^\top, \ldots, \tilde{y}_p^\top)^\top$, with each $\tilde{y}_i$ being an $n \times 1$ vector. Then,
            \begin{equation*}
                y^H P_X(\lambda_\ell) y = \tilde{y}^H \tilde{\Sigma}^X \tilde{y} = \sum_{1 \le i,j \le p} \tilde{y}_i^H \Sigma_{ij}^X \tilde{y}_j = \operatorname{Var}\left[\sum_{i=1}^p \tilde{y}_i^H X_i\right] \ge 0.
        \end{equation*}
        Therefore, $P_X(\lambda_\ell)$ is positive semi-definite. The proof is completed.
    \end{proof}

    \subsection{Proof of Theorem 3.1} \label{pf:proposedsol}
    \begin{proof}
    Denote the error to be minimized by $J$. Then, using (4), the error becomes
        \begin{equation*}
            \begin{aligned}
            J &= \sum_{i=1}^{p} E[(X_i-\hat{X}_i)^H(X_i-\hat{X}_i)] = \sum_{i=1}^{p} \operatorname{tr}[E[(X_i-\hat{X}_i)(X_i-\hat{X}_i)^H]] \\
            &= \sum_{i=1}^{p} \operatorname{tr}[E[\{(X_i-\mu_i^X -\sum_{j=1}^p \mr{G}_{ij}(Y_j-\mu_j^Y)) + (\mu_i^X-\mu_i -\sum_{j=1}^p \mr{G}_{ij}\mu_j^Y)\}  \\
            &\quad \quad \quad \quad \quad \quad \{(X_i-\mu_i^X -\sum_{j=1}^p \mr{G}_{ij}(Y_j-\mu_j^Y)) + (\mu_i^X-\mu_i -\sum_{j=1}^p \mr{G}_{ij}\mu_j^Y)\}^H]],
            \end{aligned}
        \end{equation*}
        where $Y_j$ is defined by (3), and $\mu_j^Y = E[Y_j] = \sum_{k=1}^p \mr{H}_{jk} \mu_k^X$. Note that $\sum_{j=1}^p \mr{G}_{ij}(Y_j-\mu_j^Y) = \sum_{j=1}^p \mr{G}_{ij} (\sum_{k=1}^p \mr{H}_{jk}(X_k-\mu_k^X)) = \sum_{j=1}^p \mr{A}_{ij}(X_j-\mu_j^X)$. Consequently,  

    \begin{align} \label{eq:twoterms}
        J & = \sum_{i=1}^{p} \operatorname{tr}[E[\{(X_i-\mu_i^X -\sum_{j=1}^p \mr{A}_{ij}(X_j - \mu_j^X)) + (\mu_i^X-\mu_i -\sum_{j=1}^p \mr{G}_{ij}\mu_j^Y)\} \nonumber  \\
            &\quad \quad \quad \quad \quad \quad \{(X_i-\mu_i^X - \sum_{j=1}^p \mr{A}_{ij}(X_j-\mu_j^X)) + (\mu_i^X-\mu_i - \sum_{j=1}^p \mr{G}_{ij}\mu_j^Y)\}^H]] \nonumber\\
             &= \sum_{i=1}^{p} \operatorname{tr}[\Sigma_{ii}^X - \sum_{j=1}^p \mr{A}_{ij}\Sigma_{ji}^X - \sum_{j=1}^p \Sigma_{ij}^X\mr{A}_{ij}^H + \sum_{1\le j,k \le p}\mr{A}_{ij}\Sigma_{jk}^X\mr{A}_{ik}^H]  \nonumber\\
            & \quad \quad \quad \quad \quad \quad + \; \sum_{i=1}^{p} (\mu_i^X-\mu_i -\sum_{j=1}^p \mr{G}_{ij}\mu_j^Y)^H (\mu_i^X-\mu_i -\sum_{j=1}^p \mr{G}_{ij}\mu_j^Y) \nonumber\\
            &= \sum_{\ell=1}^n \underbrace{\left[ \sum_{i=1}^{p} \left\{p_{ii}^X(\lambda_\ell) - \sum_{j=1}^p a_{ij}(\lambda_\ell)p_{ji}^X(\lambda_\ell) - \sum_{j=1}^p p_{ij}^X(\lambda_\ell)a_{ij}^*(\lambda_\ell) + \sum_{1 \le j,k \le p}a_{ij}(\lambda_\ell)p_{jk}^X(\lambda_\ell)a_{ik}^*(\lambda_\ell)  \right\}\right]}_{J(\lambda_\ell)} \nonumber\\
            & \quad \quad \quad \quad \quad \quad + \;\sum_{i=1}^{p} (\mu_i^X-\mu_i -\sum_{j=1}^p \mr{G}_{ij}\mu_j^Y))^H (\mu_i^X-\mu_i -\sum_{j=1}^p \mr{G}_{ij}\mu_j^Y)),
    \end{align}
    where $\mr{A}_{ij}=V \operatorname{diag}(\{a_{ij}(\lambda_\ell)\}_{\ell=1}^{n}) V^H$, and $\Sigma_{ij}^X$ is the cross-covariance matrix between $X_i$ and $X_j$. If we compute the solutions  $\hat{\mr{H}}_{ij}$ and $\hat{\mr{G}}_{ij}$ that minimize $J$, we obtain 
    \begin{equation*}
        \hat{\mu}_i =  \mu_i^X -\sum_{j=1}^p \hat{\mr{G}}_{ij}\mu_j^Y = \mu_i^X -\sum_{j=1}^p \hat{\mr{G}}_{ij}(\sum_{k=1}^p \hat{\mr{H}}_{jk}\hat{\mu}_k^X),
    \end{equation*} which makes the latter term on the right-hand side of (\ref{eq:twoterms}) equal to zero. 
    
    Now, let us find the solutions minimizing the former term on the right-hand side of (\ref{eq:twoterms}). To achieve this, we need to minimize $J(\lambda)$ for each $\lambda_\ell$ ($1 \le \ell \le n$). Define the matrices $A(\lambda_\ell)$, $H(\lambda_\ell)$, and $G(\lambda_\ell)$ as  $p \times p$, $q \times p$, and $p \times q$ matrices, respectively, where the $(i,j)$th elements are $a_{ij}(\lambda_\ell)$, $h_{ij}(\lambda_\ell)$, and  $g_{ij}(\lambda_\ell)$, representing the frequency responses of the graph filters $\mr{A}_{ij}$, $\mr{H}_{ij}$, and $\mr{G}_{ij}$, respectively. Then, we have $A(\lambda_\ell) = G(\lambda_\ell)H(\lambda_\ell)$, implying that $rank(A(\lambda_\ell)) \le q$. Moreover, Corollary 3.1 ensures that $P_X^{1/2}(\lambda_\ell)$ exists. With this in mind, $J(\lambda_\ell)$ becomes 
    \begin{equation*}
        \begin{aligned}
            J(\lambda_\ell)&=\operatorname{tr}[P_X(\lambda_\ell) - A(\lambda_\ell) P_X(\lambda_\ell) - P_X(\lambda_\ell)A(\lambda_\ell)^H +A(\lambda_\ell)P_X(\lambda_\ell)A(\lambda_\ell)^H]  \\
            & = \operatorname{tr}[(I-A(\lambda_\ell))P_X(\lambda_\ell)(I-A(\lambda_\ell))^H] \\
            & = \operatorname{tr}[(P_X^{1/2}(\lambda_\ell)-A(\lambda_\ell)P_X^{1/2}(\lambda_\ell))(P_X^{1/2}(\lambda_\ell)-A(\lambda_\ell)P_X^{1/2}(\lambda_\ell))^H].
        \end{aligned}
    \end{equation*} 
     Thus, from Theorem 3.7.4 of \citet{Brillinger2001time}, $J(\lambda_\ell)$ is minimized by 
     \begin{equation*}
         \hat{A}(\lambda_\ell) = \sum_{i=1}^q u_i(\lambda_\ell)u_i(\lambda_\ell)^H,
     \end{equation*}
     which is equal to $\hat{G}(\lambda_\ell) \hat{H}(\lambda_\ell)$ for $\hat{G}(\lambda_\ell) = (
					u_1(\lambda_\ell) \mid  
					\cdots \mid 
					u_q(\lambda_\ell)) = (\hat{H}(\lambda_\ell))^H$.

    For the minimizing solution discussed above, we have $I-\hat{A}(\lambda_\ell) = \sum_{i>q} u_i(\lambda_\ell)u_i(\lambda_\ell)^H$. Therefore, $J(\lambda_\ell)$ becomes $\sum_{i>q} \tau_i(\lambda_\ell)$. Consequently, the minimum mean squared error is obtained as $\sum_{\ell=1}^{n} \sum_{i>q} \tau_i(\lambda_\ell)$. This completes the proof.
    \end{proof}

    \subsection{Proof of Proposition 3.2} \label{pf:zero_coherence}
    \begin{proof}
        Denote the cross-covariance between $Y_i$ and $Y_j$ by  $\Sigma_{ij}^Y$ and the GCSD between $Y_i$ and $Y_j$ by $p_{ij}^Y$. We have 
        \begin{equation*}
            \Sigma_{ij}^Y = \operatorname{Cov}(Y_i, Y_j) = \operatorname{Cov}\left(\sum_{k=1}^p \hat{\mr{H}}_{ik}X_k, \sum_{m=1}^p \hat{\mr{H}}_{jm}X_m\right) = \sum_{k=1}^p \sum_{m=1}^p \hat{\mr{H}}_{ik} \Sigma_{km}^X \hat{\mr{H}}_{jm}^H,
        \end{equation*}
        where $\hat{\mr{H}}_{ij}$ is the matrix defined in Theorem 3.1.
        Then,
        \begin{equation*}
            p_{ij}^Y(\lambda_\ell) = \sum_{k=1}^p \sum_{m=1}^p \hat{h}_{ik}(\lambda_\ell) p_{km}^X(\lambda_\ell) \hat{h}_{jm}^*(\lambda_\ell) = u_i(\lambda_\ell)^H P_X(\lambda_\ell) u_j(\lambda_\ell) = \begin{cases}
                \tau_i(\lambda_\ell) &\text{if $i = j$,}\\
                0 &\text{if $i \neq j$.}
                \end{cases}
        \end{equation*}
        The proof is completed.    
    \end{proof}

    \subsection{Proof of Proposition 3.3} \label{pf:error_cpsd}
    \begin{proof}
    From the proof of Theorem 3.1, we know that $\epsilon_i = X_i - \mu_i^X - \sum_{j=1}^{p}\hat{\mr{A}}_{ij}(X_j - \mu_j^X)$, where $\hat{\mr{A}}_{ij}$ is the matrix defined in Theorem 3.1, so it is clear that $E[\epsilon_i] = \bs{0}$. Moreover, the cross-covariance $\Sigma_{ij}^\epsilon$ between $\epsilon_i$ and $\epsilon_j$ is obtained as
    \begin{equation*}
        \begin{aligned}
            \Sigma_{ij}^\epsilon = \operatorname{Cov}(\epsilon_i, \epsilon_j) &= \operatorname{Cov}\left(X_i - \mu_i^X - \sum_{k=1}^{p}\hat{\mr{A}}_{ik}(X_k - \mu_k^X), \, X_j - \mu_j^X - \sum_{m=1}^{p}\hat{\mr{A}}_{jm}(X_m - \mu_m^X)\right) \\
            &= \Sigma_{ij}^X - \sum_{k=1}^{p} \hat{\mr{A}}_{ik} \Sigma_{kj}^X -  \sum_{m=1}^{p} \Sigma_{im}^X \hat{\mr{A}}_{jm}^H + \sum_{1 \le k, m \le p} \hat{\mr{A}}_{ik} \Sigma_{km}^X \hat{\mr{A}}_{jm}^H.   
        \end{aligned}
    \end{equation*}
    Consequently, the GCSD $p_{ij}^\epsilon$ is expressed as
    \begin{equation*}
        p_{ij}^\epsilon(\lambda_\ell) = p_{ij}^X(\lambda_\ell) -  \sum_{k=1}^{p} \hat{a}_{ik}(\lambda_\ell) p_{kj}^X(\lambda_\ell) -  \sum_{m=1}^{p} p_{im}(\lambda_\ell) \hat{a}_{jm}^*(\lambda_\ell) + \sum_{1 \le k, m \le p} \hat{a}_{ik}(\lambda_\ell) p_{km}^X(\lambda_\ell) \hat{a}_{jm}^*(\lambda_\ell).
    \end{equation*}
    Therefore, the spectral matrix $P_\epsilon(\lambda_\ell)$ is given by 
    \begin{equation*}
        P_\epsilon(\lambda_\ell) = (I-\hat{\mr{A}}(\lambda_\ell)) P_X(\lambda_\ell) (I-\hat{\mr{A}}(\lambda_\ell))^H = \sum_{i>q} \tau_i(\lambda_\ell) u_{i}(\lambda_\ell) u_{i}(\lambda_\ell)^H.
    \end{equation*}
    \end{proof}

    \subsection{Proof of Proposition 3.4} \label{pf:xhat_coherence}
    \begin{proof}
    Given the relationships $Y_i = \sum_{l=1}^p \hat{\mr{H}}_{il}X_l$, $\epsilon_j = X_j - \mu_j^X - \sum_{m=1}^{p}\hat{\mr{A}}_{jm}(X_m - \mu_m^X)$, and  $\hat{X}_k = \mu_k^X - \sum_{s=1}^{p}\hat{\mr{A}}_{ks}(X_s - \mu_s^X)$, the covariance between $Y_i$ and $\epsilon_j$ is determined by
    \begin{equation*}
        \operatorname{Cov}\left(\sum_{l=1}^p \hat{\mr{H}}_{il}X_l, \, X_j - \mu_j^X - \sum_{m=1}^{p}\hat{\mr{A}}_{jm}(X_m - \mu_m^X)\right) = \sum_{l=1}^p  \hat{\mr{H}}_{il} \Sigma_{lj}^X - \sum_{1\le l, m \le p} \hat{\mr{H}}_{il} \Sigma_{lm}^X \hat{\mr{A}}_{jm}^H.
    \end{equation*}
    Thus, the GCSD between $Y_i$ and $\epsilon_j$ is given by 
    \begin{equation*}
        \sum_{l=1}^p  \hat{h}_{il}(\lambda_\ell) p_{lj}^X(\lambda_\ell) - \sum_{1\le l, m \le p} \hat{h}_{il} p_{lm}^X \hat{a}_{jm}^* = \left(\hat{H}(\lambda_\ell) P_X(\lambda_\ell) - \hat{H}(\lambda_\ell) P_X(\lambda_\ell) \hat{A}(\lambda_\ell)^H \right)_{i,j},
    \end{equation*}
    where $(\cdot)_{i,j}$ denotes the $(i,j)$th element of the matrix. Using the solution derived in Theorem 3.1, we can easily show that $\hat{H}(\lambda_\ell) P_X(\lambda_\ell) = \hat{H}(\lambda_\ell) P_X(\lambda_\ell) \hat{A}(\lambda_\ell)^H = \operatorname{diag}(\{\tau_i(\lambda_\ell)\}_{i=1}^{q})\hat{H}(\lambda_\ell)$. Consequently, the GCSD between $Y_i$ and $\epsilon_j$ reduces to zero. Similarly, the covariance between $\epsilon_j$ and $\hat{X}_k$ is obtained by
    \begin{equation*}
        \begin{aligned}
            \operatorname{Cov}\left(X_j - \mu_j^X - \sum_{m=1}^{p}\hat{\mr{A}}_{jm}(X_m - \mu_m^X), \, \mu_k^X + \sum_{s=1}^{p}\hat{\mr{A}}_{ks}(X_s - \mu_s^X) \right) = \sum_{s=1}^{p}  \Sigma_{js}^X \hat{\mr{A}}_{ks}^H - \sum_{1 \le m, s \le p} \hat{\mr{A}}_{jm} \Sigma_{ms}^X \hat{\mr{A}}_{ks}^H.
        \end{aligned}
    \end{equation*}
    Therefore, the GCSD between $\epsilon_j$ and $\hat{X}_k$ is given by 
    \begin{equation*}
        \sum_{s=1}^p  p_{js}^X(\lambda_\ell)\hat{a}_{ks}^*(\lambda_\ell) - \sum_{1\le m, s \le p} \hat{a}_{jm}(\lambda_\ell) p_{ms}^X(\lambda_\ell) \hat{a}_{ks}^*(\lambda_\ell) = \left(P_X(\lambda_\ell) \hat{A}(\lambda_\ell)^H - \hat{A}(\lambda_\ell) P_X(\lambda_\ell) \hat{A}(\lambda_\ell)^H \right)_{i,j}.
    \end{equation*}
    We can demonstrate that $P_X(\lambda_\ell) \hat{A}(\lambda_\ell)^H = \hat{A}(\lambda_\ell) P_X(\lambda_\ell) \hat{A}(\lambda_\ell)^H = \hat{G}(\lambda_\ell) \operatorname{diag}(\{\tau_i(\lambda_\ell)\}_{i=1}^{q}) \hat{H}(\lambda_\ell)$, thereby proving that the GCSD between $\epsilon_j$ and $\hat{X}_k$  reduces to zero. This  completes the proof.
    \end{proof}

\counterwithin{figure}{section}
    \section{Additional Figures} \label{appendix:additionalplot}
\begin{figure}[!ht]
        \centering
            \includegraphics[width=0.99\textwidth]{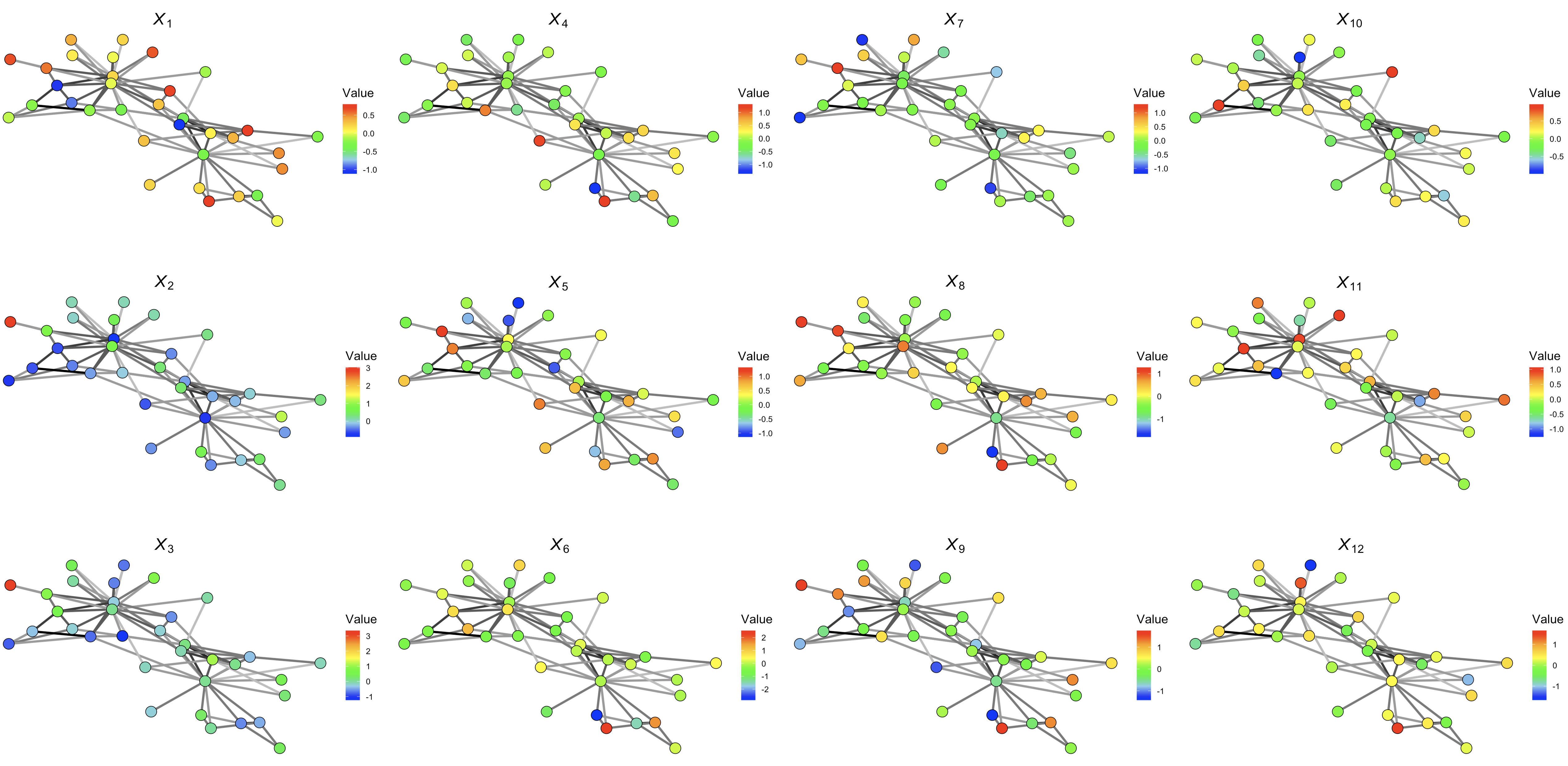}
            \vspace{-3mm}
        \caption{The 12 generated signals on the Karate club network in Section 4.1.}
        \label{fig:karate_data_plot}
    \end{figure}

\begin{figure}[!ht]
        \centering
            \includegraphics[width=0.99\textwidth]{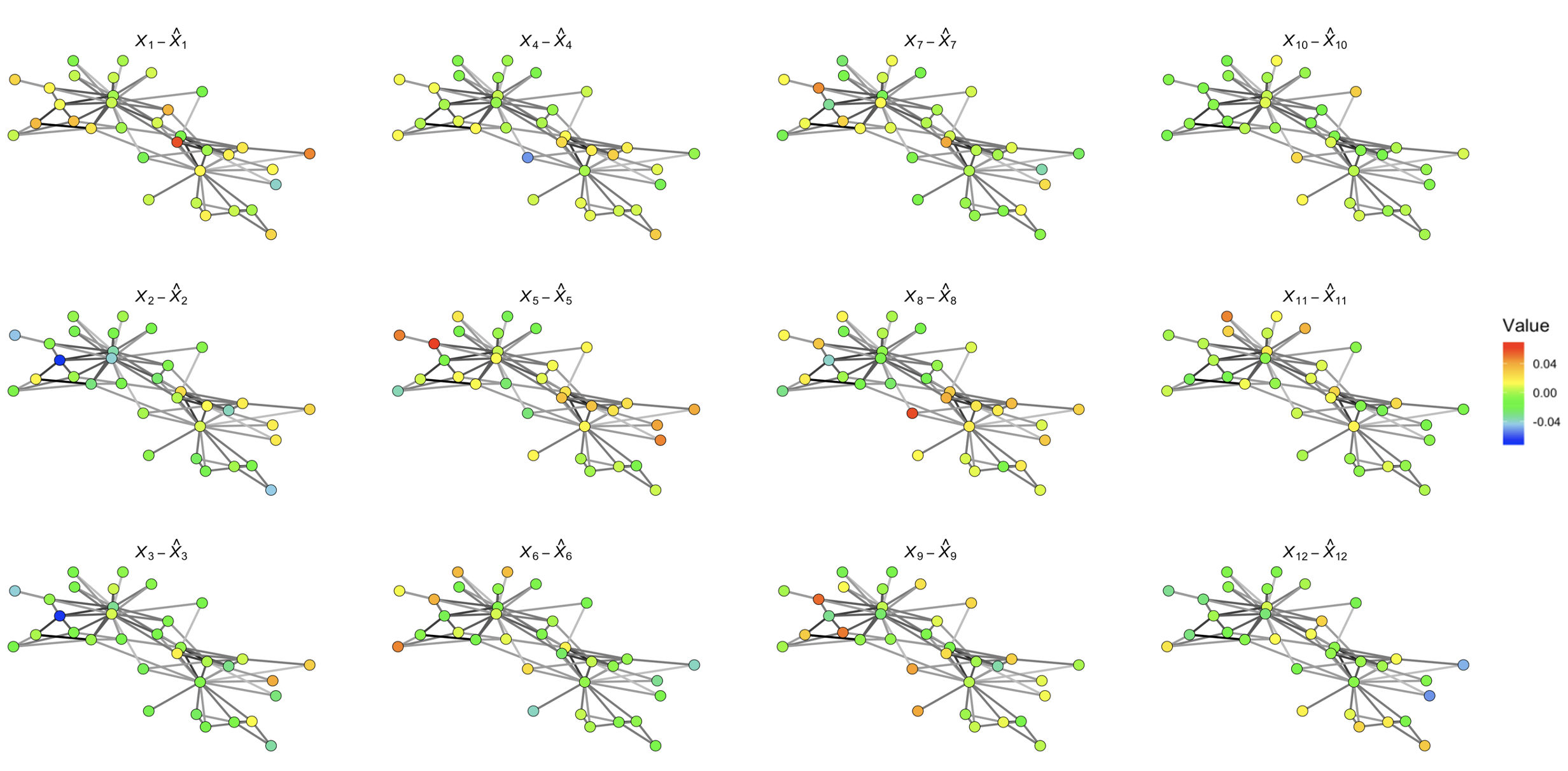}
            \vspace{-3mm}
        \caption{The 12 residual signals on the Karate club network in Section 4.1.}
        \label{fig:karate_residual_plot}
    \end{figure}

    Figures \ref{fig:karate_data_plot} and \ref{fig:karate_residual_plot} visualize the generated signals and the corresponding residual signals on the Karate club network in Section 4.1.

 \end{appendices}
	\clearpage

	\bibliographystyle{apalike}
	\bibliography{refs}

\end{document}